\documentclass[journal]{IEEEtran}

\usepackage{amsmath, amssymb, amsfonts, color}
\usepackage{graphicx, subfigure}
\usepackage{wrapfig}
\usepackage[square, numbers,sort&compress]{natbib} 
\usepackage{ulem}

\begin{document}

\title{Coherent Pattern Prediction in Swarms of Delay-Coupled Agents}

\author{Luis Mier-y-Teran-Romero,
        Eric Forgoston,
        and Ira B. Schwartz
\thanks{L. Mier-y-Teran-Romero is a joint NIH postdoctoral fellow with the
  Johns Hopkins Bloomberg School of Public Health,
615 North Wolfe Street, Baltimore, Maryland, 21205, USA and the Nonlinear Systems Dynamics Section, Plasma
  Physics Division, Code 6792, U.S. Naval Research Laboratory, Washington, DC
  20375, USA  e-mail: luis@nlschaos.nrl.navy.mil}
\thanks{E. Forgoston is with the Department of Mathematical Sciences,
  Montclair State University, 1 Normal Avenue, Montclair, NJ 07043, USA
  e-mail: eric.forgoston@montclair.edu}
\thanks{I.B. Schwartz is with the Nonlinear Systems Dynamics Section, Plasma
  Physics Division, Code 6792, U.S. Naval Research Laboratory, Washington, DC
  20375, USA e-mail: ira.schwartz@nrl.navy.mil}}

\maketitle

\begin{abstract}
We consider a general swarm model of self-propelling agents
    interacting through a pairwise potential in the presence of noise and
    communication time delay. Previous work [Phys. Rev. E 77, 035203(R)
    (2008)] has shown that a communication time delay in the swarm induces a
    pattern bifurcation that depends on the size of the coupling amplitude.
    We extend these results by completely unfolding the bifurcation structure
    of the mean field approximation. Our analysis reveals a direct correspondence between the different dynamical behaviors found in
    different regions of the coupling-time delay plane with the different classes of simulated  coherent swarm patterns.  We derive the spatio-temporal
scales of the swarm structures, and also demonstrate how the complicated
interplay of coupling strength, time delay, noise intensity, and choice of
initial conditions can affect the swarm. In particular, our studies show that
for sufficiently large values of the coupling strength and/or the time delay,
 there is a noise intensity threshold that forces a transition of the
swarm from a misaligned state into an aligned state. We show that this alignment transition
exhibits hysteresis when the noise intensity is taken to be time dependent.
\end{abstract}


\section{Introduction}
The rich dynamic behavior of interacting multi-agent, or particle, systems
has been the focus of numerous recent studies. These multi-particle systems
are capable of self-organization, as shown by the various coherent
conformations with complex structure that they generate,
even when the interactions are short range and in the absence of a leader
agent. The study of these `swarming' or `herding' systems has had many
interesting biological applications which have resulted in a better
understanding of the spatio-temporal patterns formed by bacterial colonies,
fish, birds, locusts, ants, pedestrians, etc. \cite{Budrene95, Toner95,
  Toner98,   Parrish99,Edelsteinkeshet98,Topaz04, hebling1995}. {The
  mathematical study of these swarming systems is also helpful in
  the understanding of oscillator synchronization, as in the neural phenomenon of
  central pattern generators \cite{Cohen82}}. The results of
these studies have impacted and have been successfully applied in the design of
systems of autonomous, inter-communicating robotic systems
\cite{Leonard02,Justh04, Morgan05, chuang2007}, as well as mobile sensor
networks \cite{lynch2008}.

It is possible to design swarming models for robotic motion planning,
consensus and cooperative control, and spatio-temporal formation. Pairwise potentials for individual agents can be
straightforwardly ported onto autonomous vehicles. Furthermore, these pairwise interactions can be used in conjunction
with simple scalable algorithms to achieve multi-vehicle cooperative motion
 \cite{nguyen2005}. Specific goals include: obstacle
 avoidance \cite{Morgan05}, boundary tracking \cite{hsieh2005},
 environmental sensing \cite{lynch2008,lu2011} and  decentralized  target
 tracking \cite{chung2006}.

An important problem is that of environmental consensus estimation. Here,
the individuals of the swarm communicate with each other through a network to
achieve asymptotically synchronous information about their
environment \cite{lynch2008}. Recently, consensus was extended to include time delayed
communication among agents \cite{Jad2006}. 

Task allocation is another problem of interest involving robotic swarms. The
objective is to reallocate swarm robots to perform a set of tasks in parallel
and independently of one another in an optimal way. In order to make task
reallocation more realistic it is possible to consider a time delay that
arises from the amount of time required to switch between tasks \cite{mather2011}.

Regardless of the design objective of a robotic swarm system, a comprehensive theoretical analysis of the model must
be performed in order to achieve successful algorithm design.

Many different mathematical approaches have been utilized to study aggregating
agent systems. {Some of these} studies have treated the problem at a
single-individual level, using ordinary  differential equations (ODEs) or delay differential equations (DDEs) to
describe their trajectories \cite{vicsek95,flierl99,couzin02,Justh04}. An
alternative method has been proposed by other researchers and consists of
using continuum models that consider averaged velocity and agent density fields
that satisfy partial differential equations (PDEs)
\cite{Toner95,Toner98,Edelsteinkeshet98, Topaz04}. In addition, authors also have studied the effects of noise on the swarm's behavior and have shown the existence of
noise-induced transitions \cite{Erdmann05, Forgoston08}. The study of these
systems has been enriched by tools from statistical
physics since both first and second order phase transitions have been found in
the formation of coherent states \cite{aldana07}.

An additional effect that has recently been considered is that of communication
time delays between robots. Time delay models are common in many areas of
mathematical biology including population dynamics, neural
networks, blood cell maturation, virus
dynamics and genetic networks \cite{macdonald78, macdonald89,campbell02,
  bernard04, mackey04a, mackey04b, tianjenssnepp02, jenssnepp03, monk03}. In
the context of swarming agents, it has been shown that the introduction of a
communication time delay may induce transitions between different coherent
states in a manner which depends on the coupling strength between agents
and the noise intensity \cite{Forgoston08}. {Thus far, most of the work has
  concentrated on the  case of uniform time delays among agents \cite{Kimura08}. However, the practical  engineering of
  multi-agent systems requires researchers to consider the case in which time
  delays may vary due to data processing times, problems in inter-agent
  communication, etc. The case of differing (and even time-vary\
ing)
  time delays between agents may be treated similarly to the case of a single
  delay by using a data buffer \cite{Yang10}.}

In this work, we carry out a detailed study of the bifurcation structure of
the mean field approximation used in \cite{Forgoston08} and investigate how the
bifurcations in the system are modified in the presence of noise.  Section~\ref{sec:SM} contains the swarm model, while Sec.~\ref{sec:MFA}
  contains the derivation of the mean field approximation.  The bifurcation
  analysis of the mean field equation can be found in Sec.~\ref{sec:Bif}, and
  Sec.~\ref{sec:Comp} provides a comparison of the mean field analysis with
  the nonlinear governing equations.  In Sec.~\ref{sec:Noise},
  we describe the effects of noise on the swarm, and the conclusions are
  contained in Sec.~\ref{sec:Conc}.

\section{Swarm Model}\label{sec:SM}
We consider a two-dimensional (2D) swarm that consists of $N$
identical self-propelling individuals of unit mass that
are mutually attracted  to one another in a symmetric fashion. {Hence, the
  coupling of the agents occurs via a fully connected graph.}  In
  addition, we consider the case
in which  the individuals that comprise the swarm are communicating with each
other in a stochastic environment.  Because of the finite communication times
between individuals, there is a  time delay between interactions. Assuming
  that the communication time between agents is constant and equal to {$\tau>0$}, the
swarm dynamics is  described by the following governing equations:
\begin{subequations}
\begin{align}
\dot{\mathbf{r}}_i =& \mathbf{v}_i,\label{swarm_eq_a}\\
\dot{\mathbf{v}}_i =& \left(1 - |\mathbf{v}_i|^2\right)\mathbf{v}_i -
\frac{a}{N}\mathop{\sum_{j=1}^N}_{i\neq j}(\mathbf{r}_i(t) -
\mathbf{r}_j(t-\tau)) +  \boldsymbol{\eta}_i(t),\label{swarm_eq_b}
\end{align}
\end{subequations}
for $i =1,2\ldots,N$. The terms $\mathbf{r}_i$ and
$\mathbf{v}_i$  respectively represent the 2D position and velocity of the
$i$-th agent at time $t$. The strength of the attraction is measured by the coupling
constant {$a>0$}.  The
self-propulsion and frictional drag forces on each agent is given by the
term $\left(1 -  |\mathbf{v}_i|^2\right)\mathbf{v}_i$.  Therefore, in the
absence of coupling, agents tend to move on a straight line with unit speed
$|\mathbf{v}_i| = 1$ as time goes to infinity. The term
$\boldsymbol{\eta}_i(t) = (\eta_i^{(1)}, \eta_i^{(2)})$ is a {2D} 
vector of stochastic white noise with intensity equal to $D$ such that $\langle \eta_i^{(\ell)}(t)\rangle=0$ and $\langle \eta_i^{(\ell)}(t)
\eta_j^{(k)}(t') \rangle = 2D\delta(t-t')\delta_{ij}\delta_{\ell k}$ for
$i,j=1,2,\ldots N$ and $\ell, k = 1,2$. {It is the main objective of this work to
  identify the possible swarm behaviors for different values of $a$ and $\tau$.}

The coupling between individuals arises from a time delayed, spring-like
potential. Hence, our equations of motion may be considered to be the first
term in a Taylor expansion of other more general time delayed potential
functions about an equilibrium point.

\section{Mean Field Approximation}\label{sec:MFA}

We can investigate the stability of the swarm system by deriving a mean field
  approximation of the system. The derivation involves the
  consideration of agent coordinates relative to the center of mass and the elimination of the noise
terms. The center of mass of the swarming system is given by
\begin{align}
\mathbf{R}(t) = \frac{1}{N} \sum_{i=1}^N\mathbf{r}_i(t).
\end{align}
The position of each individual  can be decomposed into
\begin{align}\label{pos_decomp}
\mathbf{r}_i = \mathbf{R} + \delta \mathbf{r}_i,  \qquad i =1,2\ldots,N,
\end{align}
where {$\delta \mathbf{r}_i$ is the vector from the center of mass to
  particle $i$ and}
\begin{align}\label{linear_dep}
\sum_{i=1}^N\delta\mathbf{r}_i(t) = 0.
\end{align}
We substitute the ansatz  given by  Eq. \eqref{pos_decomp} into the second
order system that is equivalent to Eqs. \eqref{swarm_eq_a}-\eqref{swarm_eq_b} with $D=0$.  After simplification,  one obtains
\begin{align}\label{CM1}
\ddot{\mathbf{R}} + \delta\ddot{\mathbf{r}}_i =& \left(1 - |\dot{\mathbf{R}}|^2 -
2\dot{\mathbf{R}}\cdot \delta\dot{\mathbf{r}}_i -
|\delta\dot{\mathbf{r}}_i|^2\right)(\dot{\mathbf{R}} +
\delta\dot{\mathbf{r}_i})\notag\\
& - \frac{a(N-1)}{N}\bigg(\mathbf{R}(t) - \mathbf{R}(t-\tau) +
\delta\mathbf{r}_i(t)\bigg) \notag\\
&- \frac{a}{N}\delta\mathbf{r}_i(t-\tau),
\end{align}
where we used the fact that Eq. \eqref{linear_dep} can be written as
\begin{align}
 \delta\mathbf{r}_i(t-\tau) =
-\mathop{\sum\limits_{j=1}^{N}}\limits_{i\ne j} \delta\mathbf{r}_j(t-\tau). 
\end{align}
Summing Eq. 
\eqref{CM1} over $i$ and using Eq. \eqref{linear_dep}, we find
\begin{align}\label{CM}
\ddot{\mathbf{R}}=& \left(1 - |\dot{\mathbf{R}}|^2 -
\frac{1}{N}\sum_{i=1}^N|\delta\dot{\mathbf{r}}_i|^2\right)\dot{\mathbf{R}}
\notag\\
&- \frac{1}{N}\sum_{i=1}^N\left(2\dot{\mathbf{R}}\cdot \delta\dot{\mathbf{r}}_i +
|\delta\dot{\mathbf{r}}_i|^2\right)\delta\dot{\mathbf{r}_i}   \notag\\
& -a\frac{N-1}{N}\left(\mathbf{R}(t) - \mathbf{R}(t-\tau)\right).
\end{align}

 By inserting Eq. \eqref{CM} into Eq. \eqref{CM1} it is possible to find the
 following equation for $\delta \ddot{\mathbf{r}}_i$:
\begin{align}\label{dri}
\delta\ddot{\mathbf{r}}_i=&
\left(\frac{1}{N}\sum_{j=1}^N|\delta\dot{\mathbf{r}}_j|^2 -
2\dot{\mathbf{R}}\cdot \delta\dot{\mathbf{r}}_i -  |\delta\dot{\mathbf{r}}_i|^2
\right)\dot{\mathbf{R}} \notag\\
&+ \left(1 - |\dot{\mathbf{R}}|^2 -
2\dot{\mathbf{R}}\cdot \delta\dot{\mathbf{r}}_i -
|\delta\dot{\mathbf{r}}_i|^2\right)\delta\dot{\mathbf{r}}_i\notag\\
&+\frac{1}{N}\sum_{j=1}^N\left(2\dot{\mathbf{R}}\cdot
\delta\dot{\mathbf{r}}_j + |\delta\dot{\mathbf{r}}_j|^2\right)
\ \delta\dot{\mathbf{r}}_j \notag\\
&- a \frac{N-1}{N} \delta\mathbf{r}_i - \frac{a}{N}\delta\mathbf{r}_i(t-\tau),
\end{align}
for $i =1,2\ldots,N$. 

Taken together, Eqs. \eqref{CM} and \eqref{dri} are equivalent to
Eqs. \eqref{swarm_eq_a}-\eqref{swarm_eq_b} and they merely involve a
reconstruction of the original system that is written in terms of particle coordinates
$\mathbf{r}_i$ into this new system that is written in terms of the center of
mass $\mathbf{R}$ and  coordinates relative to  the center of mass
$\delta\mathbf{r}_i$. One can see that this mapping has transformed the
original $2N$ differential equations into $2N+2$  equations.  Due to the
relation given by Eq. \eqref{linear_dep}, only   $2N$ of the transformed set of equations are independent.  Therefore, there  is no inconsistency between the original and transformed equations.

 By neglecting the fluctuation terms {$\delta \mathbf{r}_i$} from
 Eq. \eqref{CM} {and taking $N\rightarrow \infty$}, we obtain the
 following heuristic mean field approximation  for the center of mass:
\begin{align}\label{mean_field}
\ddot{\mathbf{R}}=& \left(1 - |\dot{\mathbf{R}}|^2 \right)\dot{\mathbf{R}} -a\left(\mathbf{R}(t) - \mathbf{R}(t-\tau)\right),
\end{align}
where we made the approximation $a\frac{N-1}{N}\approx a$
since we are considering the large system size limit $N\to\infty$. We will address
the validity of neglecting the fluctuation terms in Section \ref{sec:Comp}.

\section{Bifurcations in the Mean Field Equation}\label{sec:Bif}

\begin{figure}[t!]
\begin{center}
\subfigure{\includegraphics[scale=0.45]{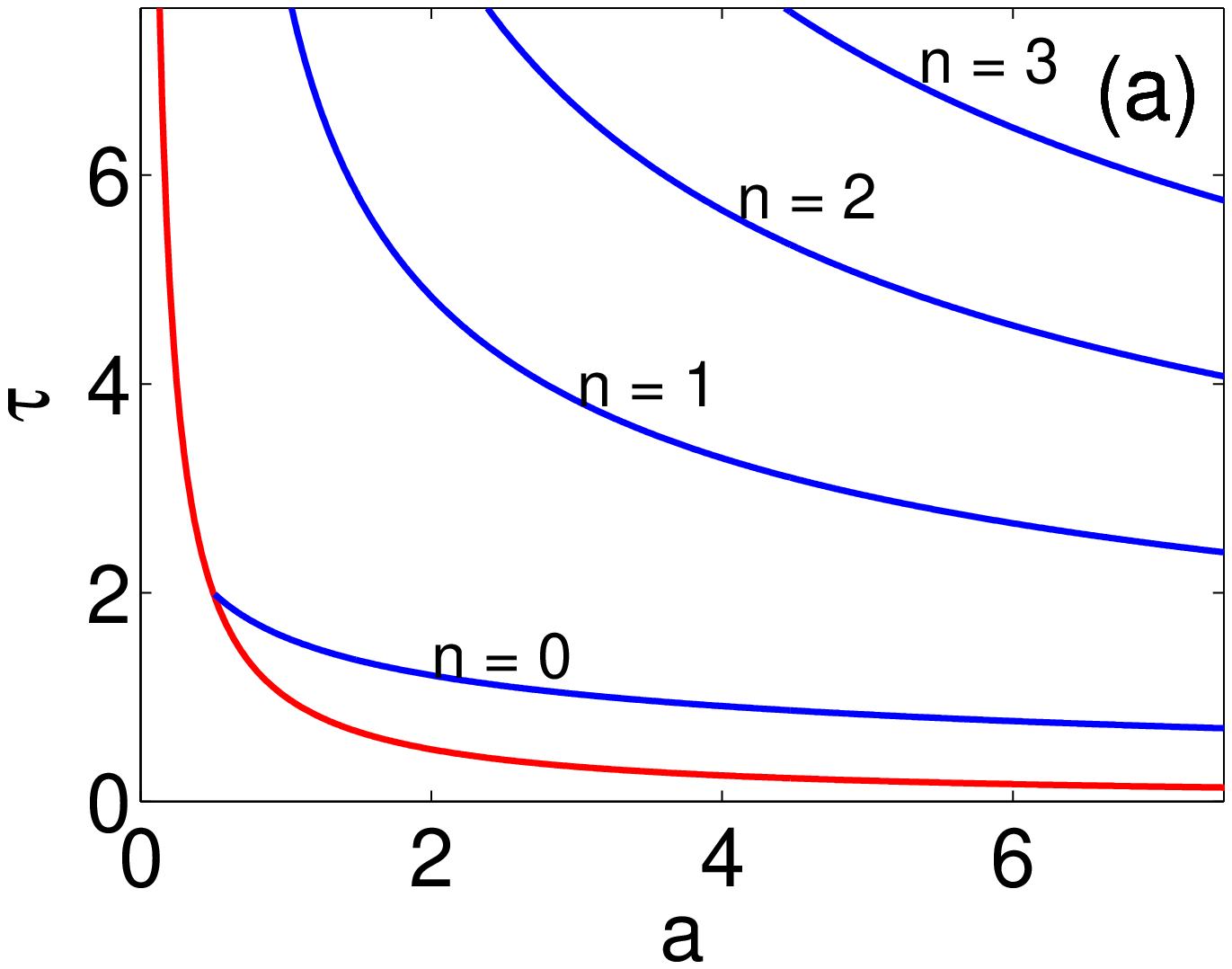} \label{Hopf_pitchfork_a_tau}}
\subfigure{\includegraphics[scale=0.45]{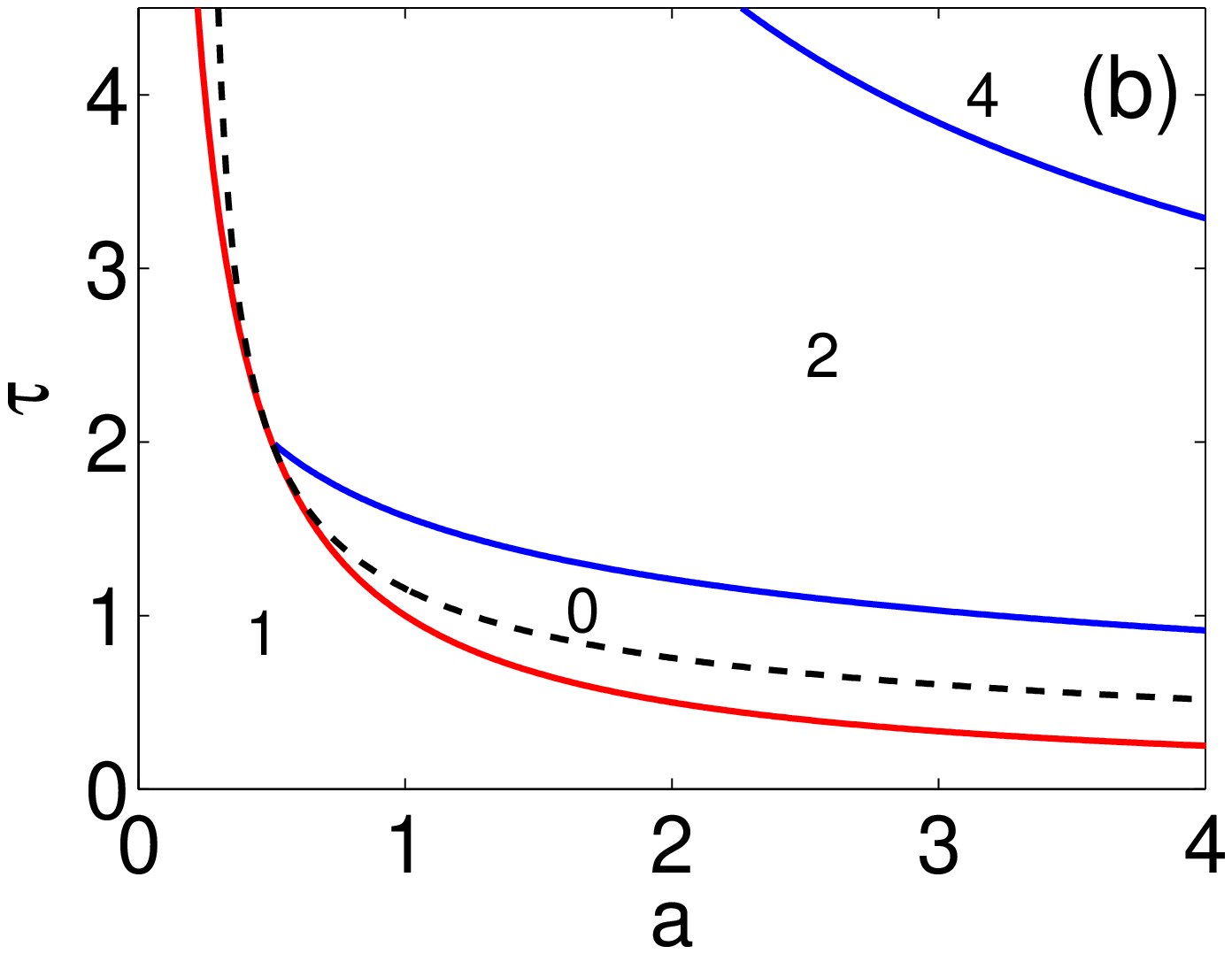} \label{Eigvls_a_tau}}
\caption{(a) Hopf (blue) and pitchfork (red) bifurcation curves in
  $(a$,$\tau)$ space. (b) A zoom-in of
  Fig. \ref{Hopf_pitchfork_a_tau}.  Included is the saddle to node
  transition curve (dashed black) and a number in each region (with boundaries given by the solid curves) that indicates the number of eigenvalues with a real part
greater than zero.}
\end{center}
\end{figure}

 Having derived a mean field equation, we continue by analyzing the
  bifurcation structure.  This bifurcation analysis will allow us to better
  understand the behavior of the system in different
regions of parameter space. {Letting}  $\mathbf{R} = (X, Y)$ and $\dot{\mathbf{R}} = (U, V)$,
{Eq. \eqref{mean_field} may be written in
component form } as 
\begin{subequations}
\begin{align}
\dot{X} &= U,\label{CM_components_a}\\
\dot{U} &= (1 - U^2 - V^2)U - a(X - X(t -\tau)),\\
\dot{Y} &= V,\\
\dot{V} &= (1 - U^2 - V^2)V - a(Y - Y(t-\tau)).\label{CM_components_d}
\end{align}
\end{subequations}
Regardless of the value of $a$ and $\tau$,
Eqs. \eqref{CM_components_a}-\eqref{CM_components_d} have translational invariant stationary solutions given by
\begin{align}
X = X_0, \quad U = 0, \quad Y = Y_0, \quad V=0,
\end{align}
where $X_0$ and $Y_0$ are two free parameters. In addition,
  Eqs. \eqref{CM_components_a}-\eqref{CM_components_d} also have
a three parameter family of uniformly translating solutions given by
\begin{align}
X = U_0 t + X_0, \quad U = U_0, \quad Y = V_0 t + Y_0, \quad V = V_0,
\end{align}
which requires
\begin{align}
 U_0^2 + V_0^2 = 1 - a\tau
\end{align}
and is real-valued only when $a\tau \le 1$. In the two-parameter space $(a, \tau)$,
the hyperbola $a \tau = 1$ is in fact a pitchfork bifurcation curve on which
the uniformly translating states are born from the stationary state $(X_0, 0, Y_0,
0)$. The pitchfork bifurcation curve can be seen in Fig. \ref{Hopf_pitchfork_a_tau}. The other branch of the pitchfork bifurcation is an unphysical solution with
negative speed.

Linearizing Eqs. \eqref{CM_components_a}-\eqref{CM_components_d} about the stationary state, we obtain the
characteristic equation
\begin{align}
\left( a(1- e^{-\lambda\tau}) - \lambda + \lambda^2 \right)^2 = 0.\label{ceq}
\end{align}
It is sufficient to study the zeros of the function 
\begin{align}\label{char_eq}
\mathcal{D}(\lambda) = a(1- e^{-\lambda\tau}) - \lambda + \lambda^2 = 0,
\end{align}
since the eigenvalues [see Eq. \eqref{ceq}] of the system given by Eqs. \eqref{CM_components_a}-\eqref{CM_components_d} are
obtained by duplicating those of Eq. \eqref{char_eq}.

We  identify the Hopf bifurcations in the two parameter space $(a,
\tau)$ by letting the eigenvalue be purely imaginary.  Our choice
  of $\lambda = i \omega$ is substituted into Eq. 
\eqref{char_eq}, {and one obtains}
\begin{align}\label{hopf_cond}
a - \omega^2 - i\omega = a e^{-i\omega \tau}.
\end{align}
By taking the modulus of Eq. \eqref{hopf_cond}, one finds
  that $a$ at the Hopf point  is given by
\begin{align}
a_H^2 = (a_H - \omega^2)^2 + \omega^2.
\end{align}
If we consider the case when $\omega \neq 0$,  then
\begin{align}\label{a_H}
a_H = \frac{1 + \omega^2}{2}.
\end{align}

We substitute Eq. \eqref{a_H} into Eq. \eqref{hopf_cond} and take the complex conjugate.  This allows us to obtain  the following equation
for $\tau$ at the Hopf point that does not involve $a$:
\begin{align}
\frac{1 - \omega^2}{1+\omega^2} + i\frac{2\omega}{1 + \omega^2} =  e^{i\omega \tau}.
\end{align}
We isolate $\tau$ by equating the arguments of both sides, being careful to
use the branch of $\tan\theta$ in $(0,\pi)$ since the left hand side of the
 equation above is on the upper complex plane for $\omega > 0$. We then
obtain a family of Hopf bifurcation curves parameterized by $\omega$:
\begin{subequations}
\begin{align}
a_H(\omega) &= \frac{1 + \omega^2}{2},\label{hopf_omega_a}\\
\tau_{H_n}(\omega) &=
\frac{1}{\omega}\left(\arctan\left(\frac{2\omega}{1-\omega^2}\right) +
2n\pi\right), \quad n = 0, 1,\ldots\label{hopf_omega_b}
\end{align}
\end{subequations}
The first few members of the family of Hopf bifurcation curves are shown in
Fig. \ref{Hopf_pitchfork_a_tau}.  It  also is possible to eliminate the
parameter $\omega$ in  Eqs. \eqref{hopf_omega_a}-\eqref{hopf_omega_b}.  Doing so, one obtains

\begin{align}\label{hopf_a}
&\tau_{H_n}(a) =\notag\\
&\frac{1}{\sqrt{2a -1}}\left(\arctan\left(\frac{\sqrt{2a-1}}{1-a}\right) +
2n\pi\right), \quad n = 0, 1,\ldots
\end{align}

In spite of their appearance, the Hopf curves in Eqs. \eqref{hopf_omega_a}-\eqref{hopf_omega_b} and
\eqref{hopf_a} are in fact continuous at $\omega = 1$ and $a=1$,
respectively [with the correct branch of $\tan\theta$ in $(0, \pi)$].
  Inspection of Eq. \eqref{hopf_omega_a}, shows that the Hopf
frequency depends only on the value of $a$ for all members in the
family.  The frequency equals one when $a=1$, and the frequency tends to
infinity as $a$ increases. Interestingly, only the first Hopf curve is
defined at $a=1/2$ and has the value $\tau_{H_0}\vert_{a=1/2} = 2$. The point ($a=1/2$, $\tau = 2$) which lies both on the first Hopf curve and on  the pitchfork  curve is a Bogdanov-Takens (BT) point  (the
  eigenvalues are zero), where $\omega = 0$. None of the other Hopf branches
meet the pitchfork bifurcation curve since $\tau\to\infty$ as $a\to 1/2$.

\begin{figure}[t!]
\begin{center}
\subfigure{\includegraphics[scale=0.26]{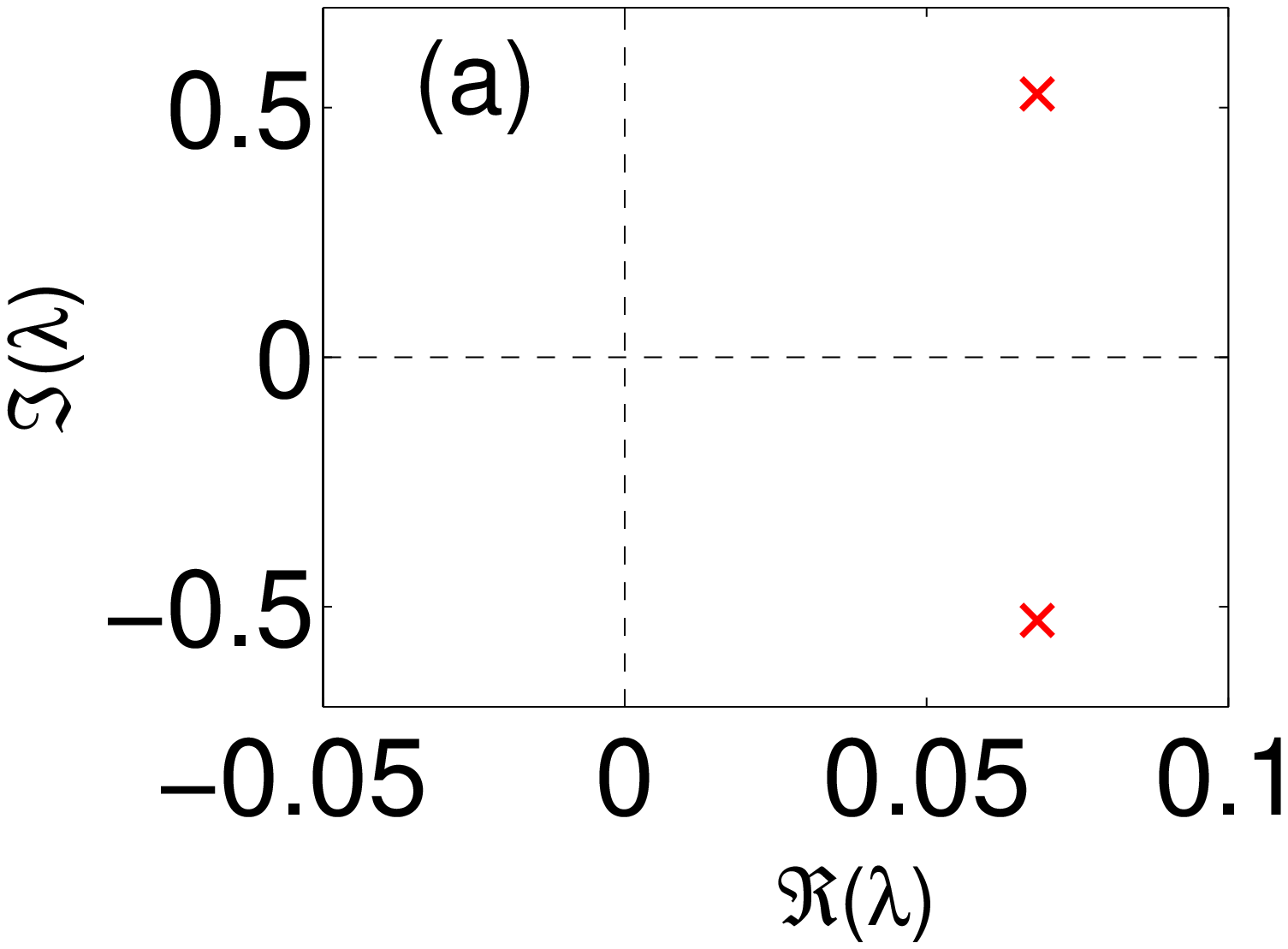} \label{BT_a}}
\subfigure{\includegraphics[scale=0.26]{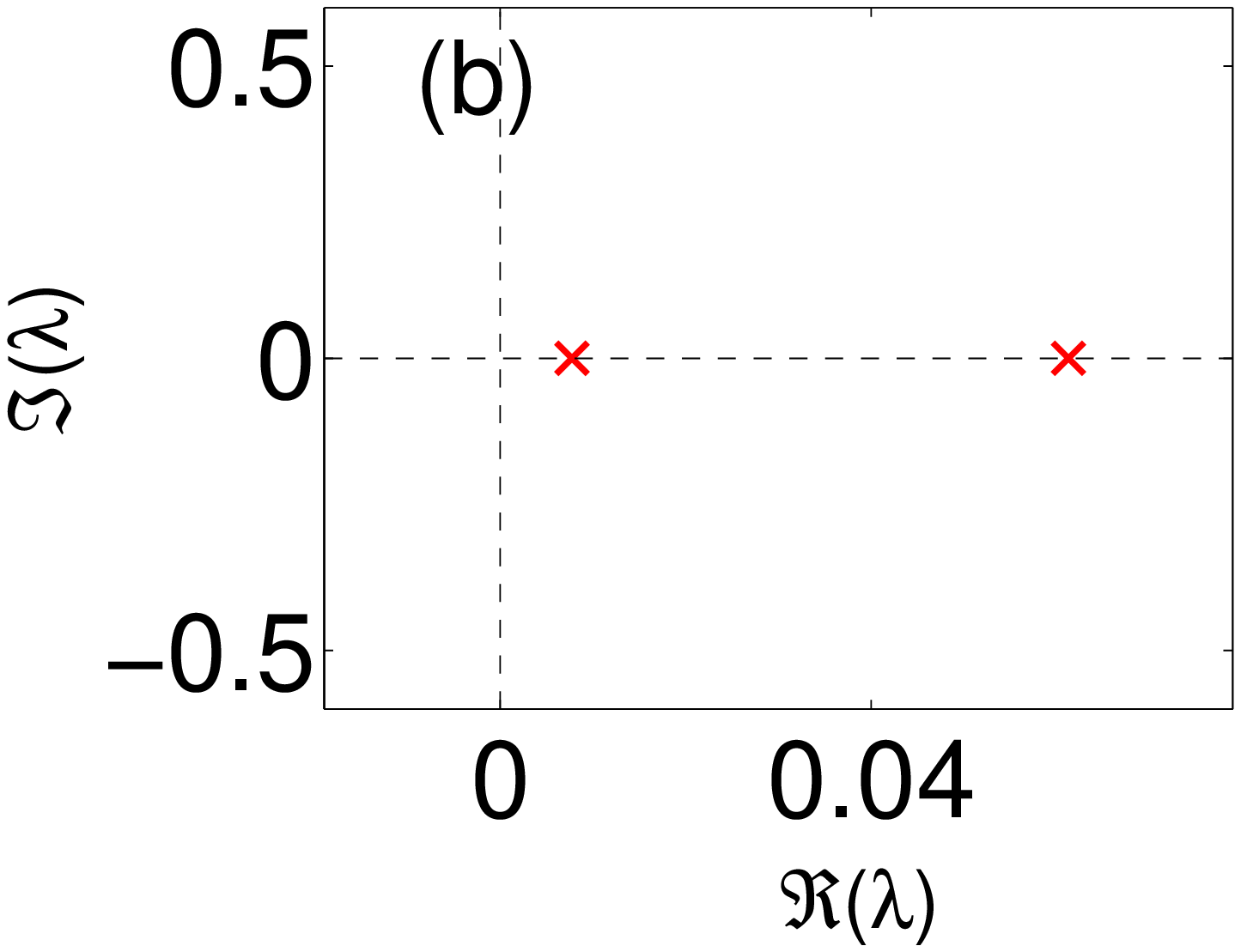} \label{BT_b}}
\subfigure{\includegraphics[scale=0.26]{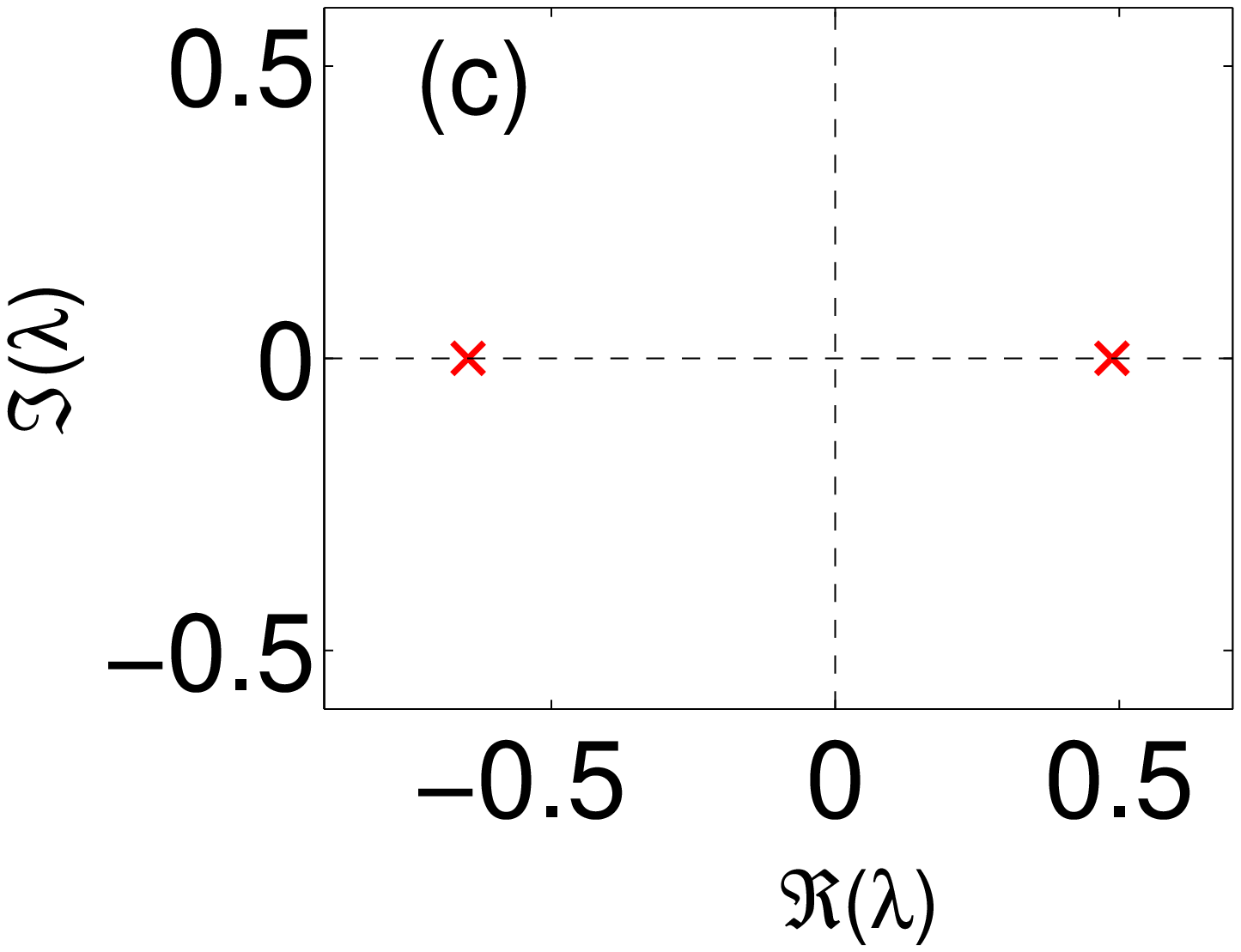} \label{BT_c}}
\subfigure{\includegraphics[scale=0.26]{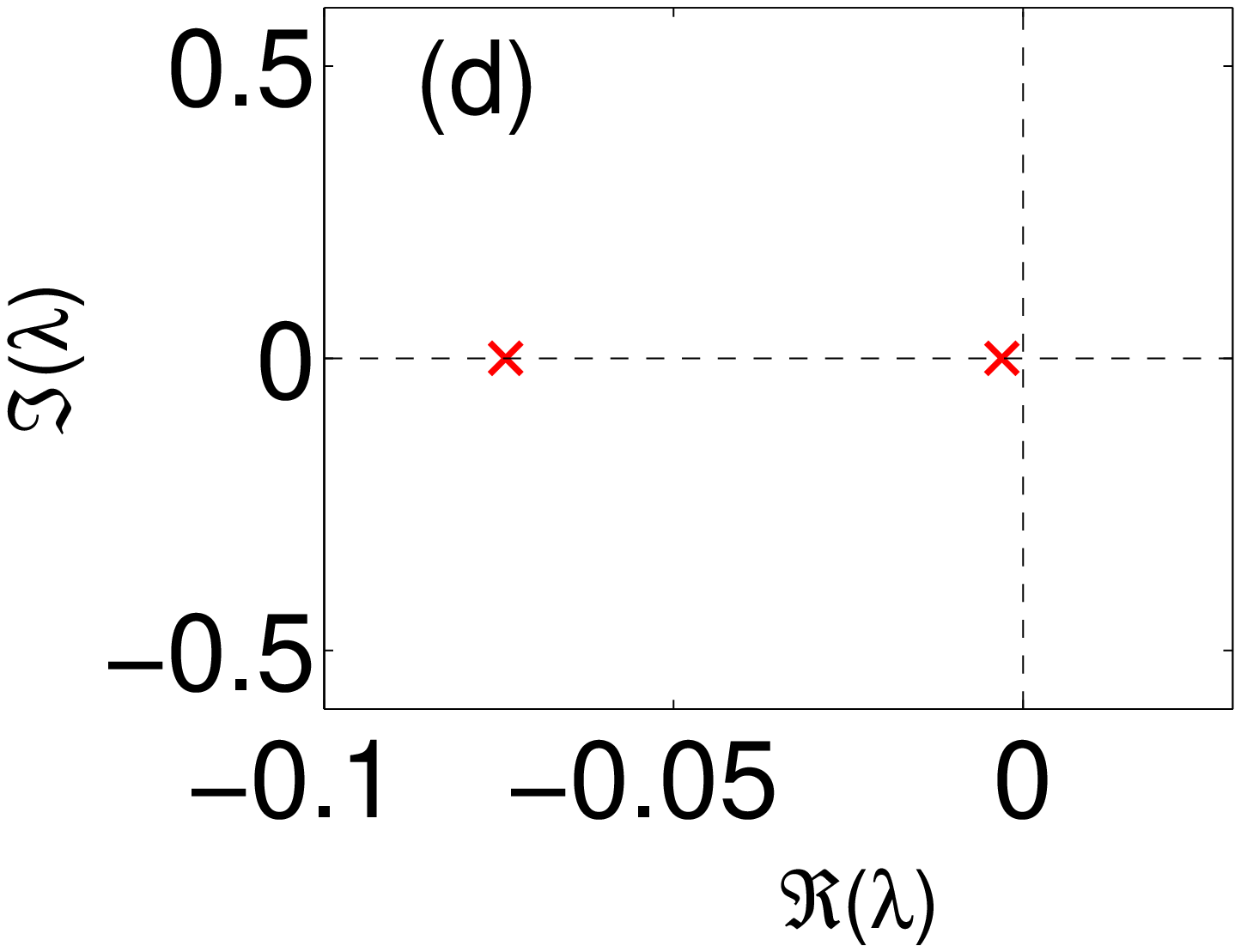} \label{BT_d}}
\subfigure{\includegraphics[scale=0.26]{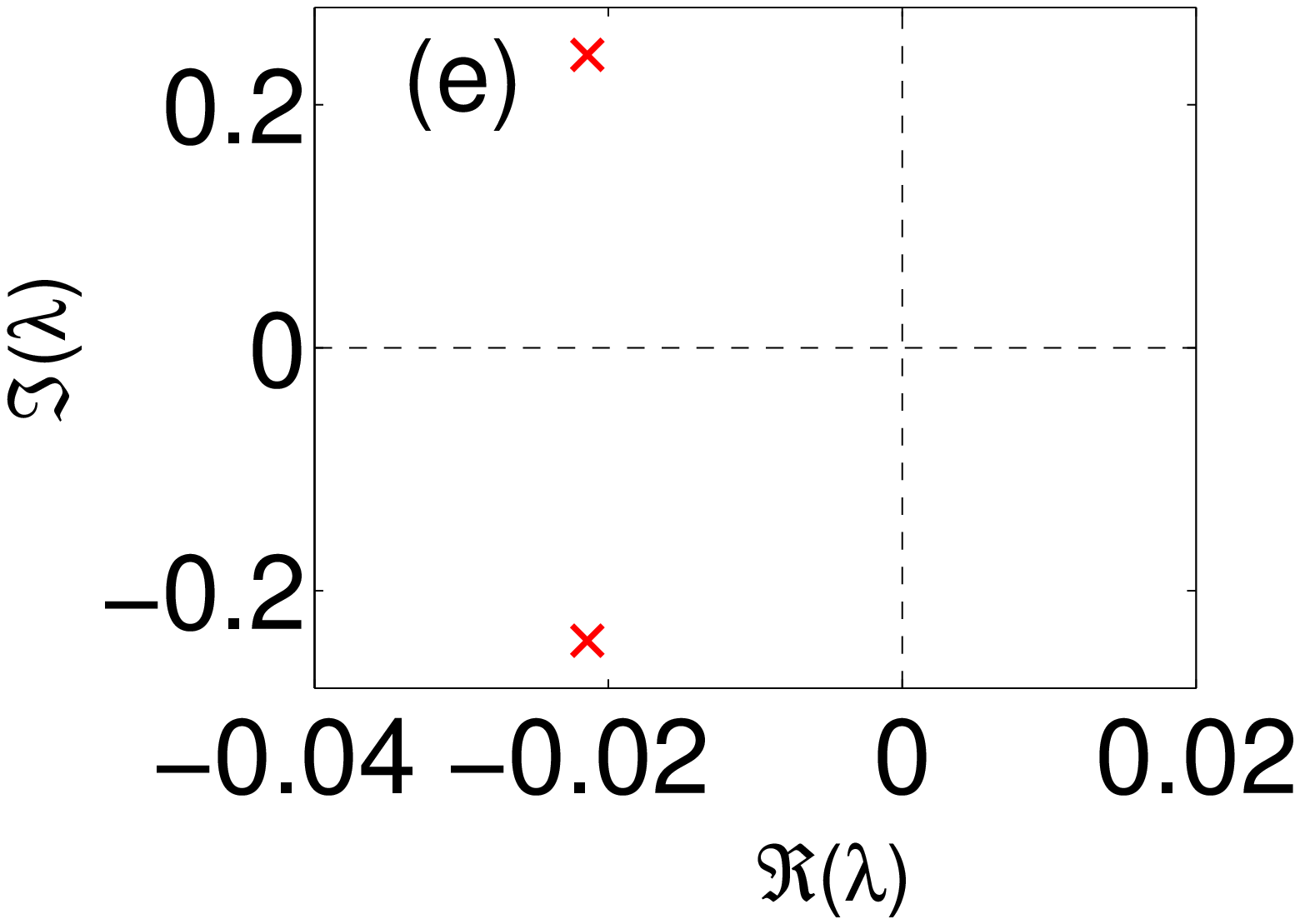} \label{BT_e}}
\caption{Real and Imaginary parts of the dominating eigenvalues as one
  moves around the Bogdanov-Takens point $(a = 1/2, \tau = 2)$ in
  $(a ,\tau)$ parameter space.  The eigenvalues shown are associated
with the locations (a) $a = 0.60$, $\tau  = 2.0$, (b) $a = 0.48$,
$\tau  = 2.09$, (c) $a = 0.40$,   $\tau  = 2.01$, (d) $a = 0.53$, $\tau  = 1.90$, and  (e) $a = 0.55$, $\tau  = 1.91$.  Refer to Fig. \ref{Eigvls_a_tau} to see where each of the $(a, \tau)$ points lies in relation to the bifurcation curves.}\label{BT_fig}
\end{center}
\end{figure}

The pitchfork and Hopf bifurcation curves in the $(a,
\tau)$ parameter space  were computed using a numerical continuation
  method \cite{Engel}.  These
results {(not shown)} are in perfect agreement with our analytical calculations.  These numerical continuation studies also allow for the determination
of the number of eigenvalues with real part greater than zero in different regions of the $(a,\tau)$ parameter space.  The results are shown in Fig. \ref{Eigvls_a_tau}. In
addition, our numerical continuation analysis revealed node to focus transitions of the steady
state. These transitions occur at points where there are two real and equal
eigenvalues, i.e. where $\mathcal{D}(\lambda) = 0$ and $\mathcal{D}'(\lambda)
= 0$, for real-valued $\lambda$. If $\mathcal{D}'(\lambda) = 0$  then one can show that $e^{-\tau\lambda} =\frac{1-2\lambda}{a\tau}$.  Insertion of this relation into $\mathcal{D}(\lambda) = 0$ leads to
\begin{align}\label{node_focus_quadratic}
\lambda^2 - \left(1 - \frac{2}{\tau}\right)\lambda + a - \frac{1}{\tau} = 0,
\end{align}
which has solutions $\lambda = \frac{1}{2}\left[1 - \frac{2}{\tau} \pm \sqrt{ 1
    +\frac{4}{\tau^2} -    4a}\right]$. For the roots to be repeated, we
set the discriminant equal to zero and this gives the following curve where the node-focus transitions occur:
\begin{align}\label{node_focus}
\tau = \frac{1}{\sqrt{a - 1/4}}.
\end{align}
Moreover, by inspecting  the solutions to Eq. \eqref{node_focus_quadratic} one finds that the repeated eigenvalues
have positive real parts if $\tau > 2$ and negative real parts if $\tau <
2$. In Figure \ref{Eigvls_a_tau}, we
show the pitchfork and Hopf bifurcation curves overlaid with the node-focus
transition curve given by Eq. \eqref{node_focus}.

As seen in Fig. \ref{Eigvls_a_tau}, the pitchfork and first Hopf
bifurcation curves, together with the node-focus transition curve, split the
area around the BT point into five different regions. The behavior of the
dominating eigenvalues (excluding the one at the origin) in each of these five
regions is shown in Figs. \ref{BT_a}-\ref{BT_e}. Starting at a point directly to
the right of the BT point in $(a, \tau)$ space, there is a pair of
eigenvalues with positive real parts and non-zero imaginary parts
[Fig. \ref{BT_a}]. Moving counter-clockwise, the eigenvalue pair collapse on
the  positive real axis upon crossing the
upper branch of the node-focus transition  curve [Fig. \ref{BT_b}]. Continuing in the same direction, we observe two different instances of eigenvalues
crossing the origin: (i) first the smaller of the two purely real and positive
eigenvalues does so as the upper part of the  pitchfork
bifurcation  curve is crossed [Fig. \ref{BT_c}] and (ii) then the remaining
purely real and positive eigenvalue crosses the origin as the lower part of
the pitchfork bifurcation  curve is crossed [Fig. \ref{BT_d}]. Finally, as the
node-focus transition curve is crossed, the two purely
real and negative eigenvalues coincide on the negative real axis and acquire
non-zero imaginary parts [Fig. \ref{BT_e}]. Continuing upwards in parameter
space, the complex pair of eigenvalues crosses the imaginary axis
as the Hopf bifurcation curve is crossed, giving birth to a stable limit cycle.

\section{Comparison of the Mean Field Analysis and the Full Swarm Equations}\label{sec:Comp}

\begin{figure}[t!]
\begin{center}
\includegraphics[scale=0.45]{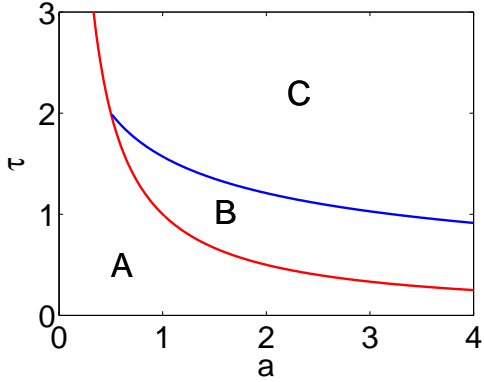} 
\caption{Regions in $(a, \tau)$ space with different dynamical behavior.}\label{a_tau_regions}
\end{center}
\end{figure}

Our analysis of the deterministic mean field equations identified the different dynamical
behaviors that the approximation given by Eq. \eqref{mean_field} exhibits in different regions of
the $(a, \ \tau)$ plane.  However,  the  analysis does not provide any information about how the
swarm agents are distributed about the center of mass. We neglect the stochastic terms in Eqs. \eqref{swarm_eq_a}-\eqref{swarm_eq_b}
and use extensive numerical simulations to identify some of the coherent structures that the swarm adopts
asymptotically in time:
\begin{itemize}
\item[(i)] A translational state, in which all swarm particles have identical
  positions and velocities and move uniformly in a straight line. The direction of motion depends on
  the initial conditions. This behavior is only possible in region A of
  Fig. \ref{a_tau_regions}. {Moreover, the asymptotic convergence to this
    state requires that all particles be located in close proximity and with
    aligned velocities at the initial time. Hence, the basin of attraction
    is extremely small which causes this state to be very sensitive
    to perturbations. This is discussed in more detail below.}

\item[(ii)] A ring state, in which the center of mass is stationary.
  The swarm agents distribute themselves along the ring with roughly half
  of the agents moving clockwise and half  of the agents moving counter-clockwise. The final stationary
  position  of the center of mass and the particular  behavior of each
    individual in the swarm is dependent on the initial conditions. This behavior is
  possible in regions A, B and C of Fig. \ref{a_tau_regions}.

\item[(iii)] A rotational state, in which all swarm agents collapse
  to the center of mass and the latter rotates on a circular orbit. The
  direction of rotation depends on the initial conditions. This behavior is only possible in region C of  Fig. \ref{a_tau_regions}.

\item[(iv)] A degenerate rotational state, in which all swarm particles collapse
  to the center of mass and the latter oscillates back and forth on a
  line. This behavior is only possible in region C of
  Fig. \ref{a_tau_regions}. In addition, it requires that the initial motion of all swarm
  particles be constrained to a line and so is sensitive with respect to
  perturbations and noise.
\end{itemize}

The above list is not extensive and our simulations have revealed other time-asymptotic
patterns. However, all of these other patterns (and including the
translational state and the degenerate rotational state) require extreme
symmetry in the initial conditions and are very sensitive with respect to
perturbations and noise. Our numerical simulations suggest that only the ring and the rotational
state have a significant robustness with respect to perturbations and noise.

The full system of equations predict a bistable behavior since the translating
and ring states are both possible in region A and C
[Fig. \ref{a_tau_regions}], depending on the initial conditions. The
linear stability analysis of Section \ref{sec:Bif} shows that the mean field approximation fails to capture this bistable behavior.

{The mean-field bifurcation results obtained here are of practical value
  since they provide us with guidelines
  for selecting values for $a$ and $\tau$ that will result in a 
  particular coherent pattern asymptotically in time. In the case of
  bistability,} our numerical simulations strongly suggest that the initial alignment of the
agents' velocities is critical in determining the coherent state
adopted. Specifically, to obtain the translating, rotating and degenerate
rotating states asymptotically in time (structures in which the individuals'
velocities are perfectly aligned), one requires a high alignment of the
initial particles' velocities; otherwise, the swarm will adopt the ring
state. However, how high an alignment is needed depends on the specific choice of $(a,\tau)$.  Our results indicate that it is easier to obtain aligned states
for larger values of the coupling constant $a$. {Unfortunately, it is not 
feasible to obtain analytic basin boundaries in this infinite dimensional
 system. In principle, one may approximate such boundaries by performing
 prohibitively extensive
 numerical simulations where the space of history functions is restricted in
 some way. Therefore, the computation of basins of attraction is outside the
 scope of this work and is left for future research.}

For the non-degenerate and degenerate rotating states as well as for the
translating state, the approximation we made when neglecting the fluctuation
terms in Eq. \eqref{mean_field} is entirely valid since in the noiseless case all agents collapse to the center of
mass. In the case of the ring structure, these fluctuation terms are
not necessarily small. However, in Eq. \eqref{CM} all fluctuation terms with the
exception of the one containing the factor $\frac{1}{N}\sum_{i=1}^N |\delta
\dot{\mathbf{r}}_i|^2$ approximately cancel out in the long time limit, due to
the symmetry in the distribution of the agents. The fluctuation term that
remains becomes equal to one in the long time limit.  This has the effect of
eliminating the self-propulsion of the center of mass and what remains is solely
cubic dissipation.

The following sub-sections contain
  detailed discussion regarding the spatio-temporal scales of
each coherent structure.

\subsection{The Ring State}

The  analysis of Appendix \ref{ring} shows that the radius and angular
frequency of the swarm particles on the ring state is given by
\begin{gather}\label{ring_rho_omega}
\rho_j = \frac{1}{\sqrt{a}}, \qquad \dot{\theta}_j =  \pm\sqrt{a},
\end{gather}
so that particles move at unit speed, $\rho_j \dot\theta_j = \pm 1$.

\begin{figure}[t!]
\begin{center}
\subfigure{\includegraphics[scale=0.26]{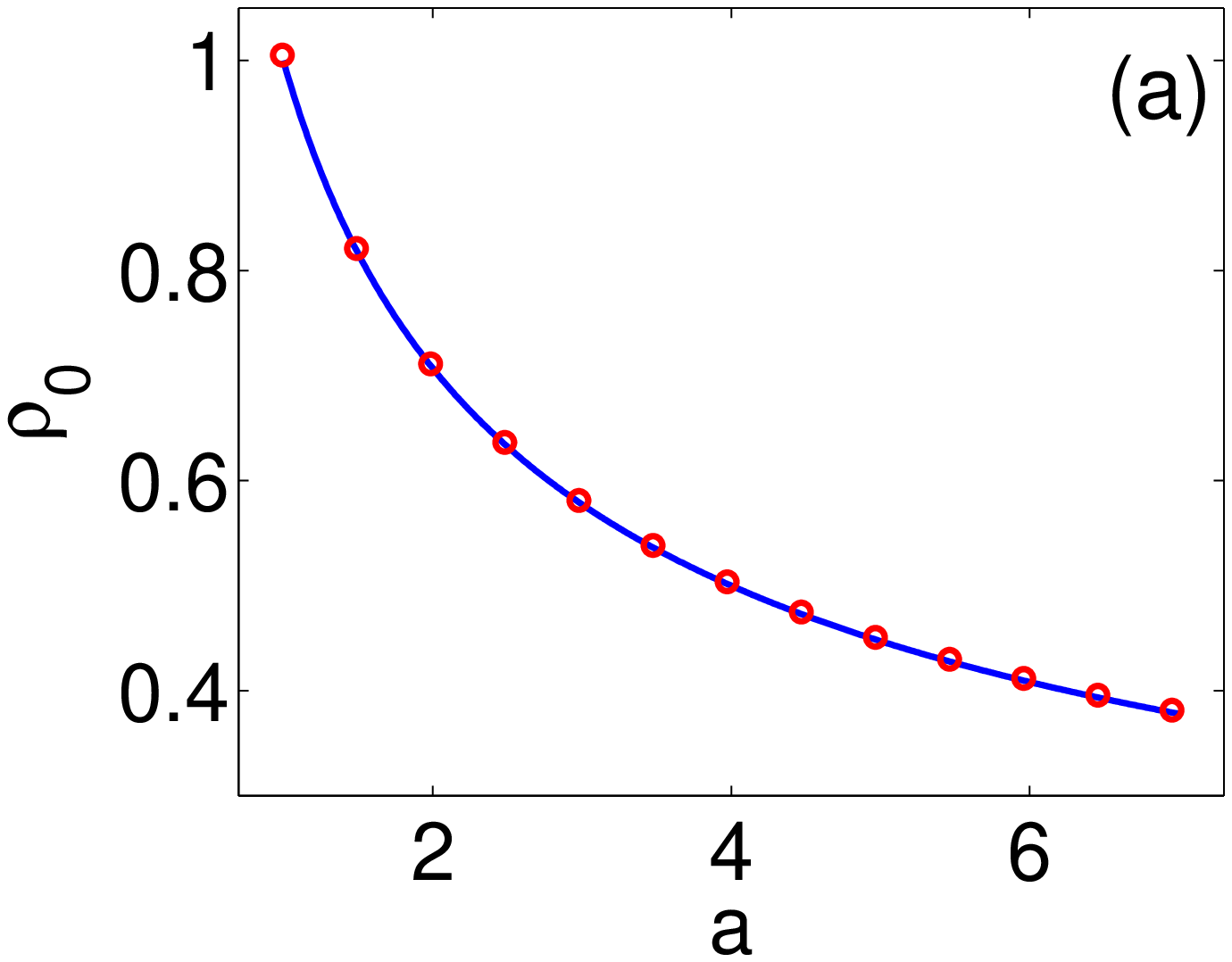} \label{Ring_radius}}
\subfigure{\includegraphics[scale=0.26]{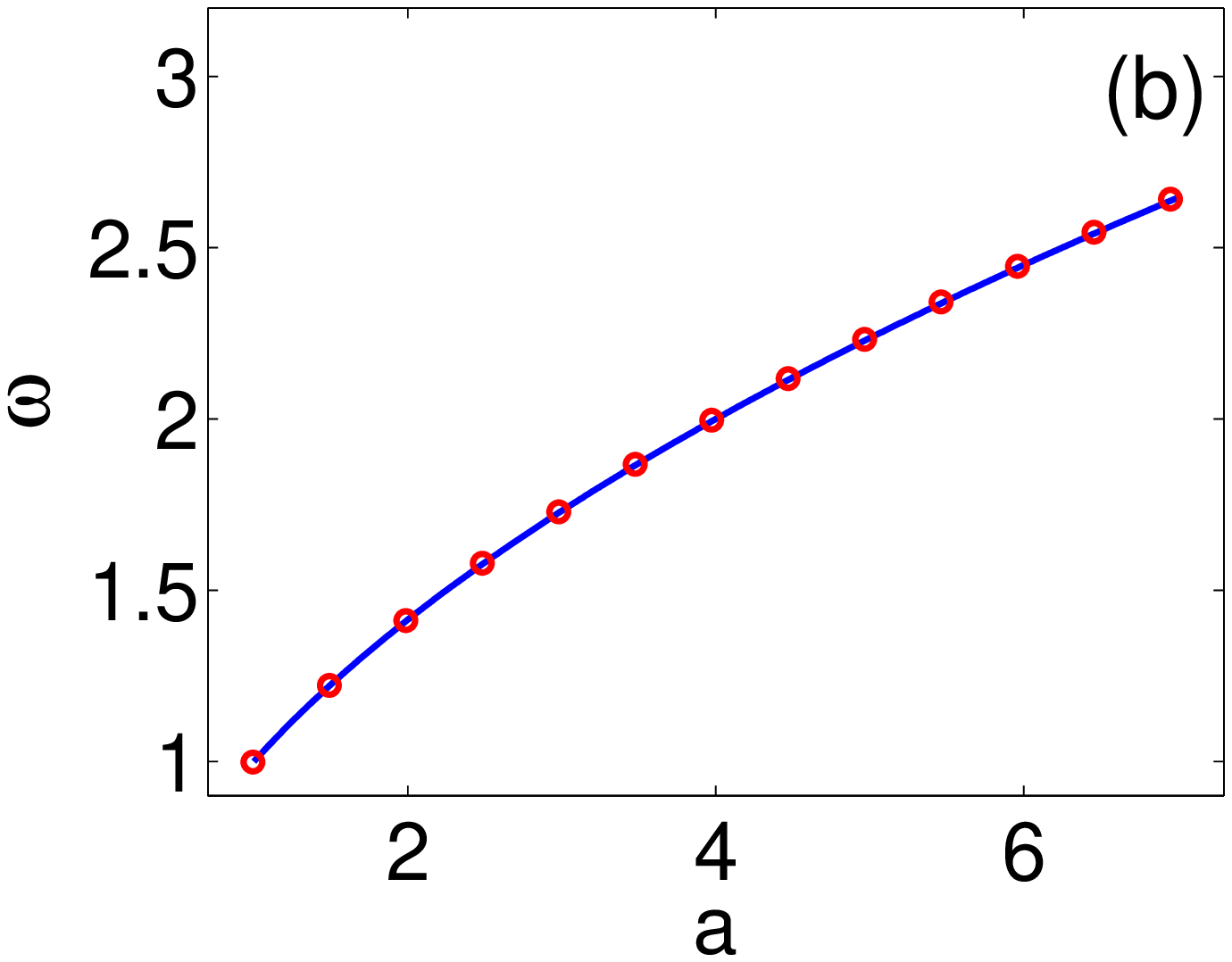} \label{Ring_omega}}
\subfigure{\includegraphics[scale=0.26]{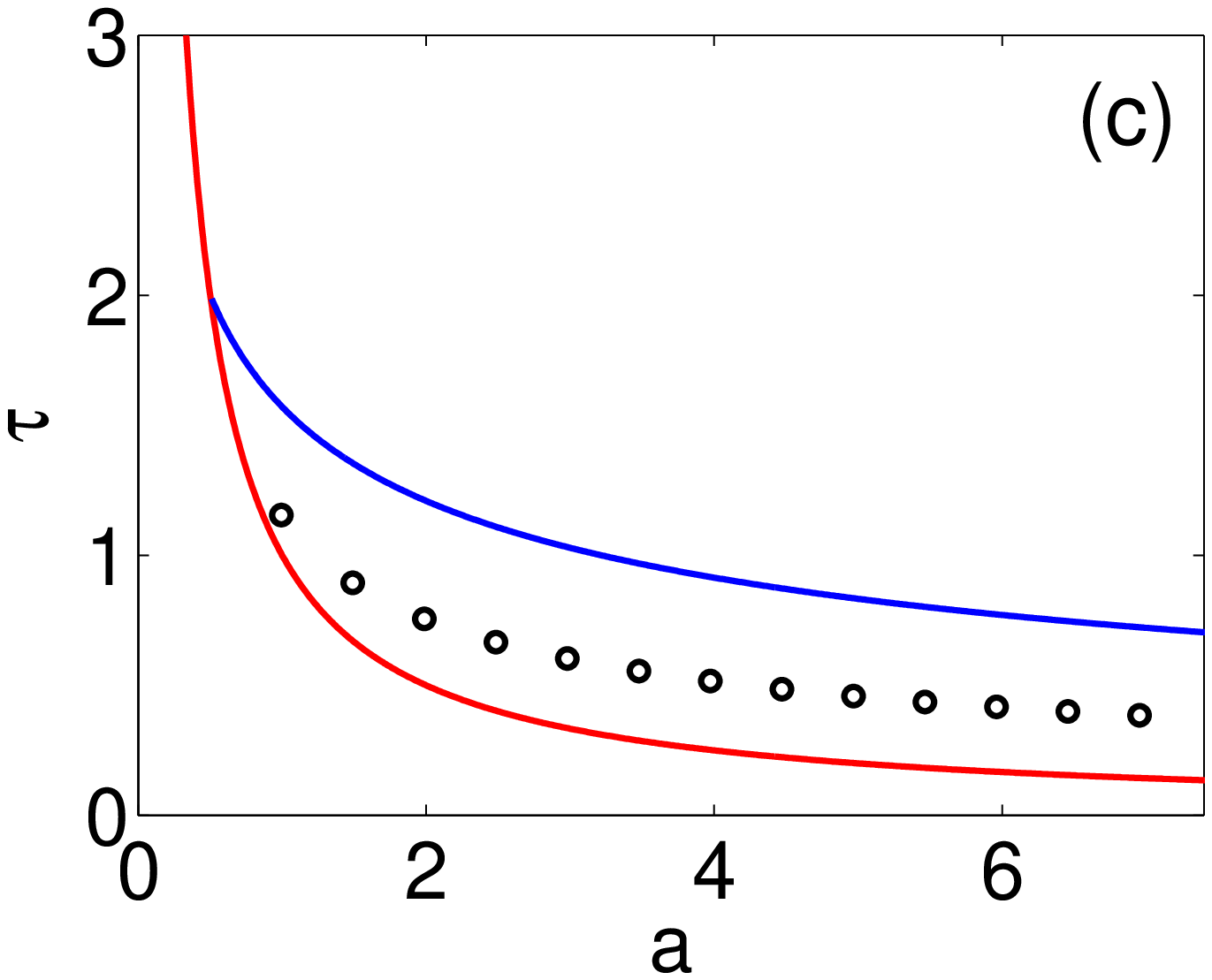} \label{Ring_sim_map}}
\caption{Comparison of numerical simulations (red
  circular markers) with the  analytical expressions (continuous blue curve) given by
    Eq. \eqref{ring_rho_omega} for (a) the radius  and (b) the frequency  of the ring  state.(c) For each value of
  $a$, the time delay was chosen as $\tau = 1/\sqrt{a-1/4}$ (black
  circular markers).}\label{Ring_fig}
\end{center}
\end{figure}

 We have numerically computed the radius and angular frequency for different values of $a$ and $\tau$
within the region in which the mean field approximation gives a stable
stationary center of mass (Fig. \ref{Ring_fig}).   Figures
  \ref{Ring_radius}-\ref{Ring_omega}  shows that there is excellent agreement
  between the numerical simulations and the analytical result given by
  Eq. \eqref{ring_rho_omega}.  It is worth noting that the condition given by Eq. \eqref{dri_condition} and used to derive Eq. \eqref{ring_rho_omega} is satisfied in the long time limit in our simulations.

\subsection{The Rotating State}

We show in Appendix \ref{rotating} that the circular orbit of the rotating state has radius
$\rho_0$ and frequency $\omega$ that satisfy the following relations:
\begin{subequations}\label{omega_rho_CM_circle}
\begin{align}
\omega^2 =& a \cdot(1 - \cos\omega\tau),\label{omega_CM_circle}\\
\rho_0 =& \frac{1}{|\omega|} \sqrt{1 - a\frac{\sin\omega\tau}{\omega}}\label{rho_CM_circle}.
\end{align}
\end{subequations}

\begin{figure}[t!]
\begin{center}
\includegraphics[scale=0.45]{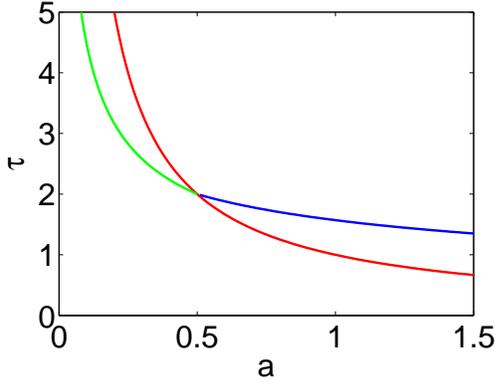} 
\caption{In $(a$,$\tau)$ space, we plot: Hopf (blue) and pitchfork (red)
  bifurcation curves, and the curve $a \tau^2=2$ where the first limit cycle
  ceases to exist by having its radius diverging to infinity (green).}\label{Circle_birth_death}
\end{center}
\end{figure}

Eqs. \eqref{omega_CM_circle}-\eqref{rho_CM_circle} can have as many solutions as desired by
choosing $a$ and $\tau$ large enough. However, a careful analysis  reveals
that the solutions to Eqs. \eqref{omega_CM_circle}-\eqref{rho_CM_circle} are
generated exactly along the Hopf curves of our previous mean field analysis and represent the same limit cycles of
that analysis [Fig. \ref{Hopf_pitchfork_a_tau}]. The expressions in
Eqs. \eqref{omega_CM_circle}-\eqref{rho_CM_circle} thus determine the
spatio-temporal scales of these circular orbits beyond the Hopf curves where
they are born. Our analysis also shows that
the circular limit cycle that is created on the first member of the Hopf bifurcation curves persists to the left of the pitchfork bifurcation curve
and then ceases to exist as its radius diverges to infinity on the curve
$a\tau^2=2$ (Fig. \ref{Circle_birth_death}). Moreover, numerical simulations
of the mean field equations reveal that both the translating state and the
rotating state are linearly stable for $(a,  \tau)$  pairs inside the wedge between the curve
$a\tau^2=2$ and the pitchfork bifurcation curve $a\tau=1$ above the BT point.

\begin{figure}[t!]
\begin{center}
\subfigure{\includegraphics[scale=0.26]{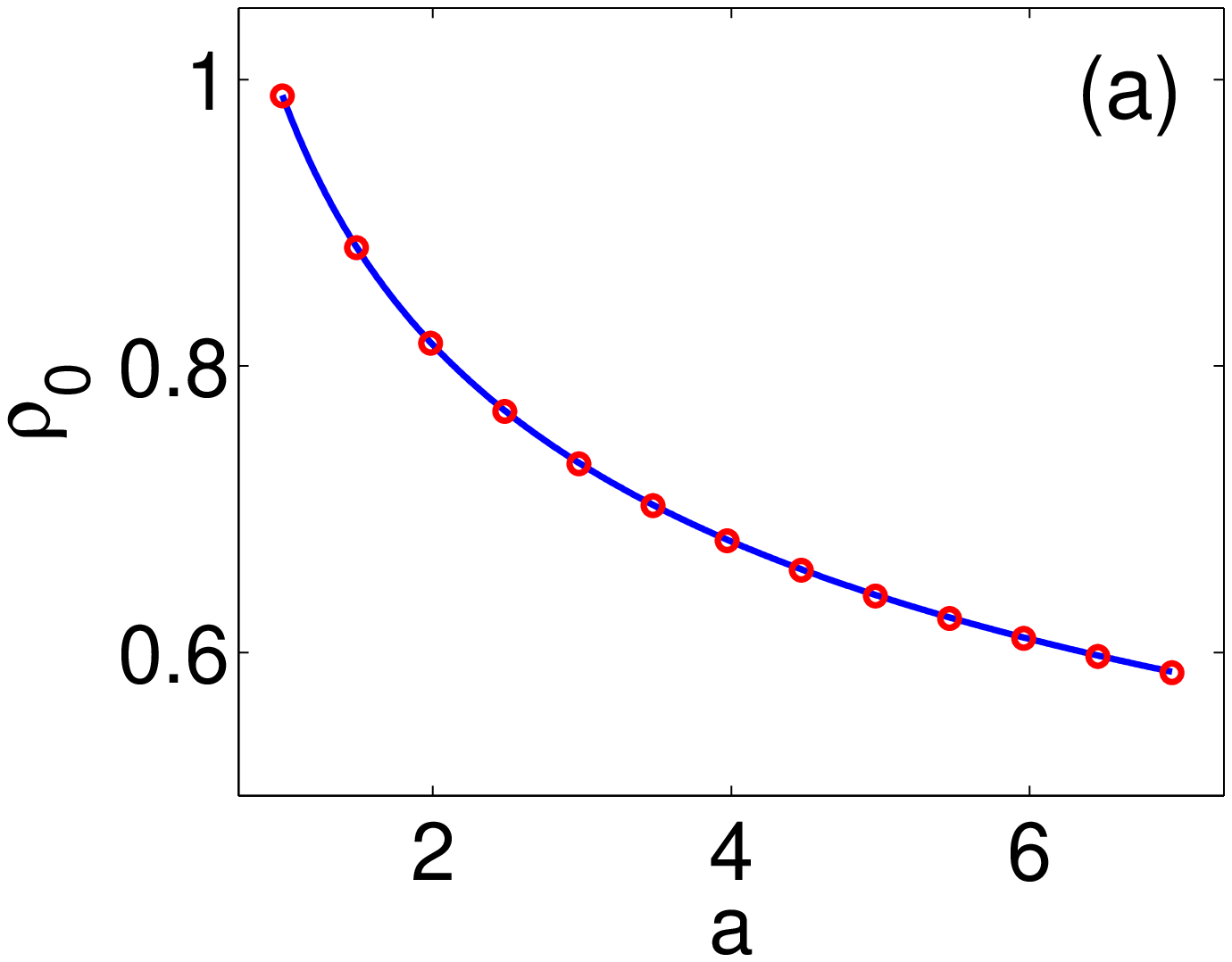} \label{Circle_radius}}
\subfigure{\includegraphics[scale=0.26]{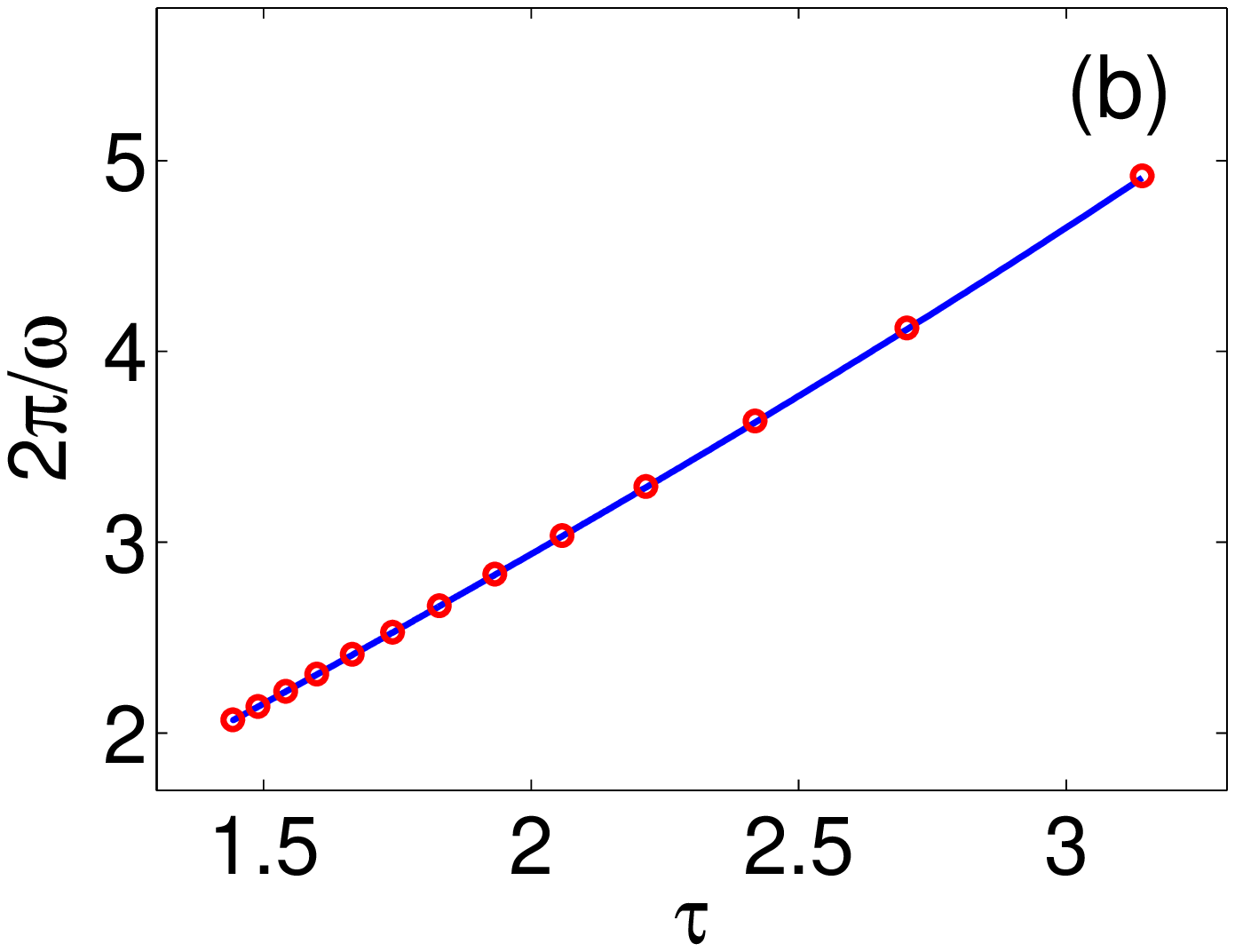} \label{Circle_omega}}
\subfigure{\includegraphics[scale=0.26]{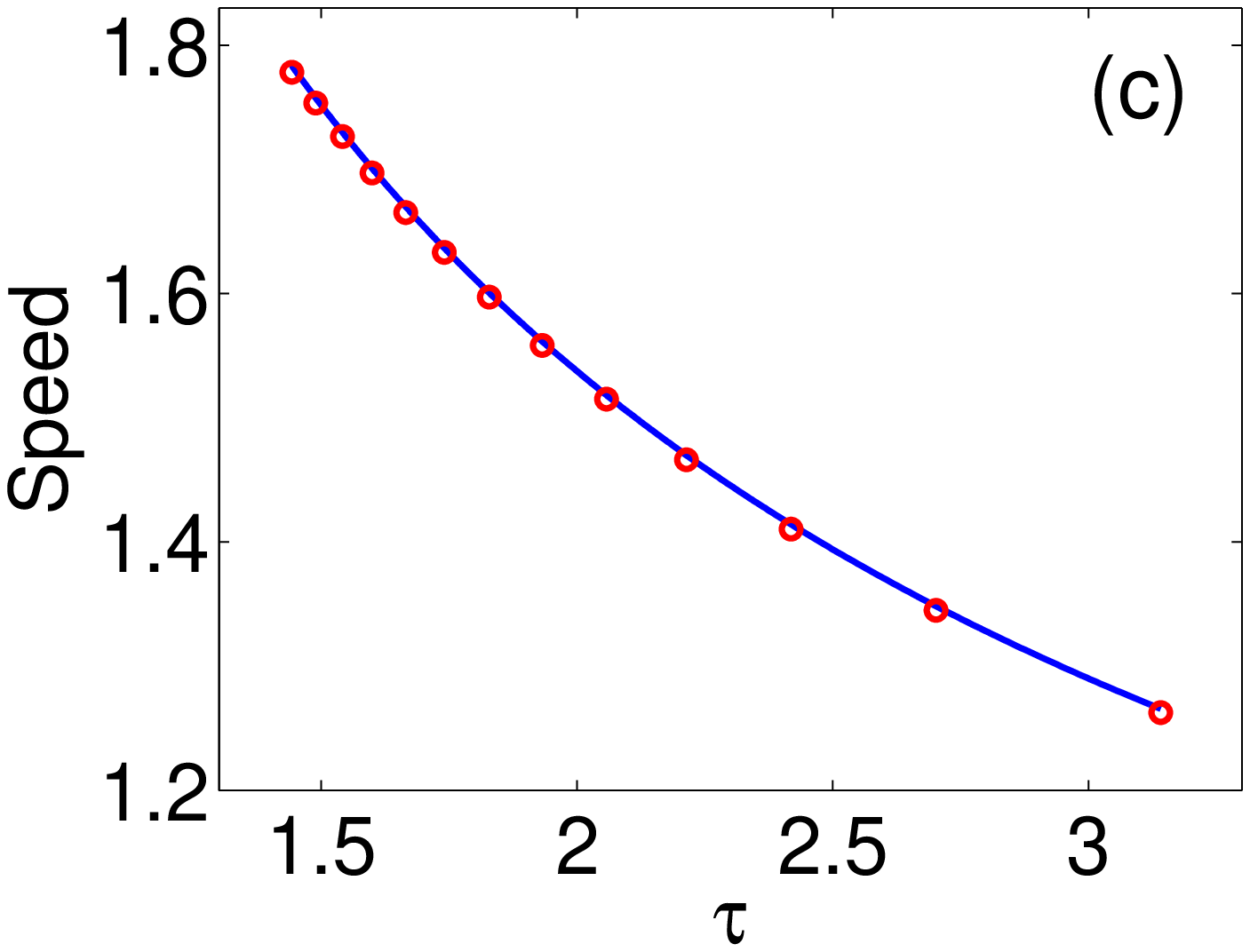} \label{Circle_speed}}
\subfigure{\includegraphics[scale=0.26]{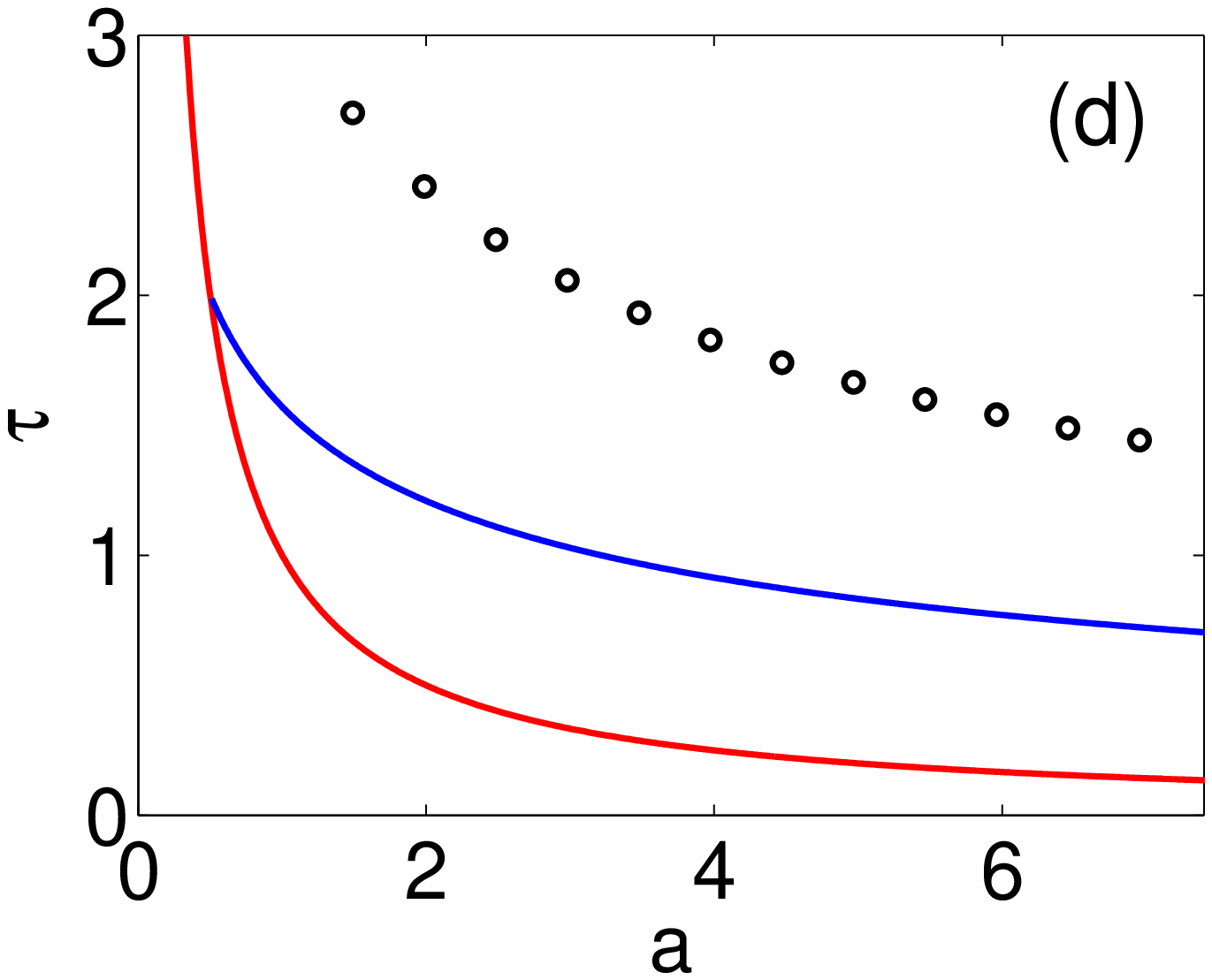} \label{Circle_sim_map}}
\caption{Comparison of  numerical simulations (red
  circular markers)  with the  analytical expressions (continuous blue curve)  given by   Eqs. \eqref{omega_CM_circle}-\eqref{rho_CM_circle} for  (a) the
    radius, (b) the period,  and  (c) the speed
   of the collapsed circular orbit.  (d) For each value of
  $a$, the time delay was chosen as $\tau =
\frac{2}{\sqrt{2a
    -1}}\arctan\left(\frac{\sqrt{2a-1}}{1-a}\right)$ (black
  circular markers)  to
  assure asymptotic time convergence to the collapsed circular orbit state.}
\label{Circle_orbit}
\end{center}
\end{figure}

Figures \ref{Circle_radius}-\ref{Circle_sim_map} show the excellent  agreement between numerical
simulations and the analytical results  given by Eqs. \eqref{omega_CM_circle}-\eqref{rho_CM_circle}, for
different values of $a$ and $\tau$.

 Interestingly, in Fig. \ref{Circle_speed} we note that in the
asymptotic time limit the collapsed agents move at a speed greater than
one, the speed at which agents would tend to move in the absence of
coupling. This is explained by noting that the ratio of the time delay to the
period of oscillations is such that the delayed position of the collapsed
agents $\mathbf{R}(t-\tau)$ is ahead of the present position
$\mathbf{R}(t)$. The attraction that an individual particle feels to the
delayed position of the rest of the swarm forces the whole system go faster.

\subsection{The Degenerate Rotating State}

A degenerate version of the rotating state is possible when the initial motion
of the swarm is restricted to a line, since in this case it follows from
Eqs. \eqref{swarm_eq_a}-\eqref{swarm_eq_b} that the swarm will remain on such
a line for all times. As we show in Appendix \ref{deg_rotating}, we may assume that the motion of the
collapsed swarm occurs on the $X=Y$
line of the center of mass coordinates and then use a finite Fourier mode
approximation of the ensuing dynamics. An
approximation in terms of just three modes gives
\begin{align}\label{X_CM_line}
X(t)=Y(t) = 2 c_1 \cos\omega t + 2|c_3| \cos(3\omega t + \phi_3), 
\end{align}
where $\omega$, $c_1$, $c_3$ and $\phi_3$ are obtained by solving
Eqs. \eqref{13_fourier_a}-\eqref{13_fourier_b} numerically.

\begin{figure}[t!]
\begin{center}
\subfigure{\includegraphics[scale=0.26]{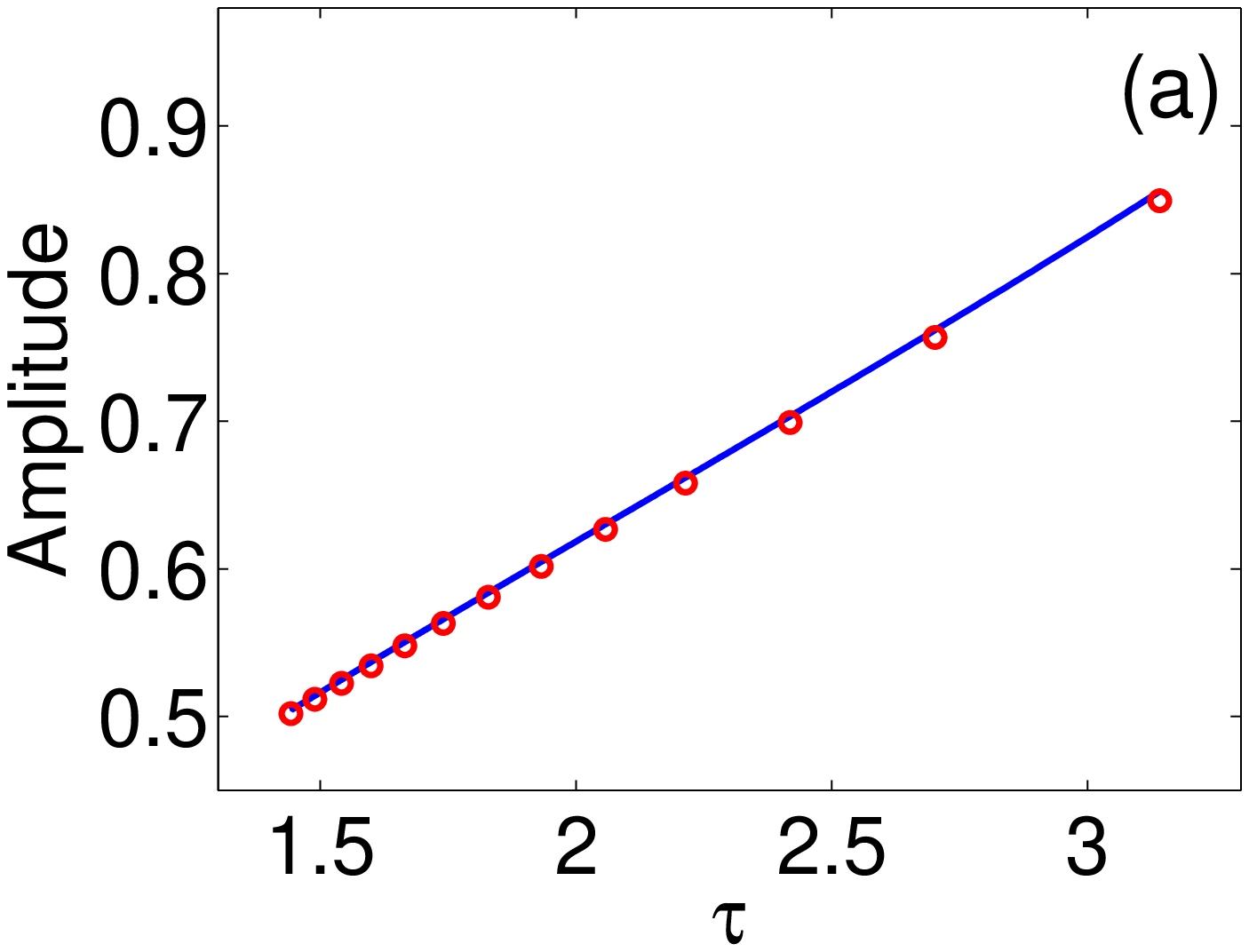} \label{Line_amplitude}}
\subfigure{\includegraphics[scale=0.26]{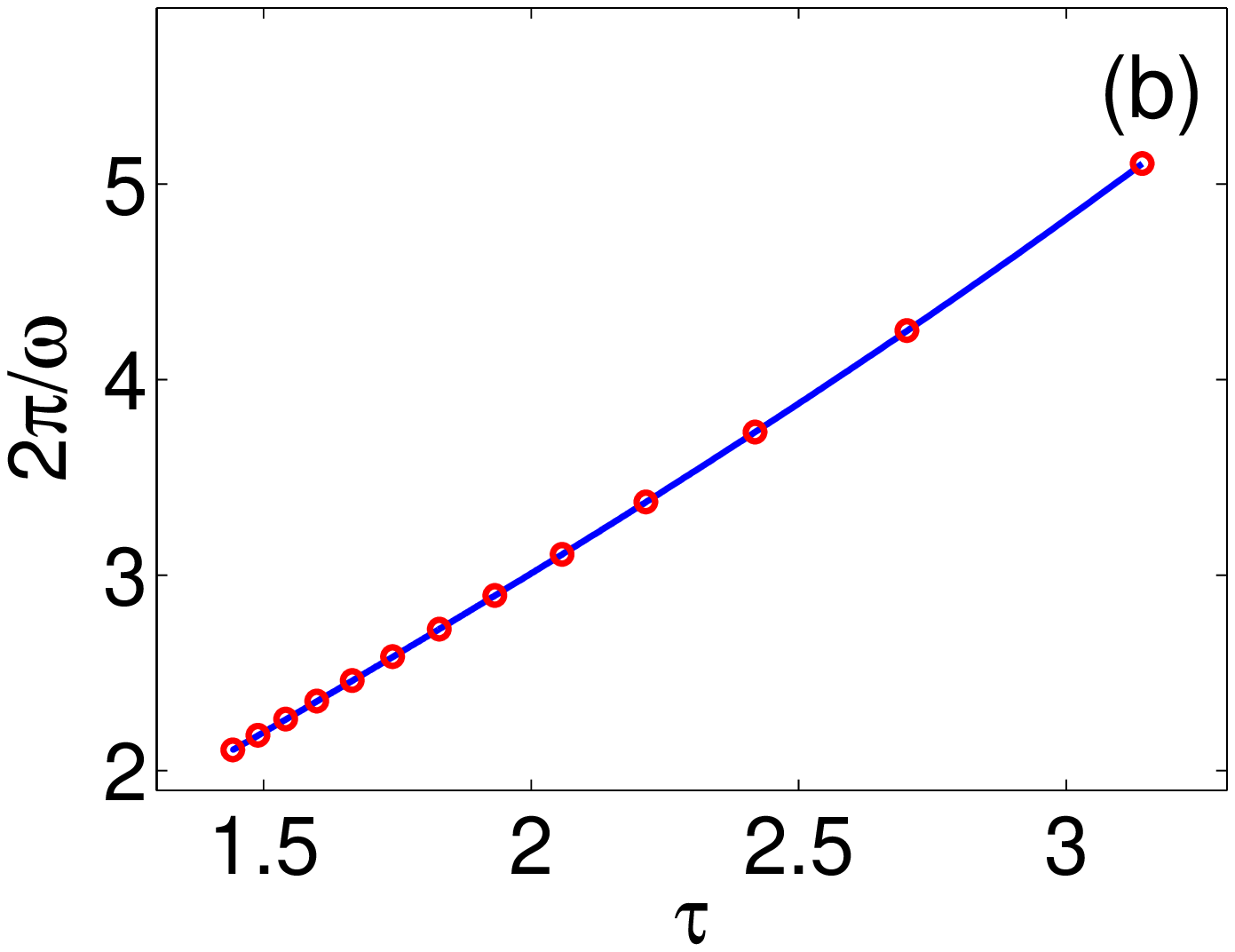} \label{Line_omega}}
\subfigure{\includegraphics[scale=0.26]{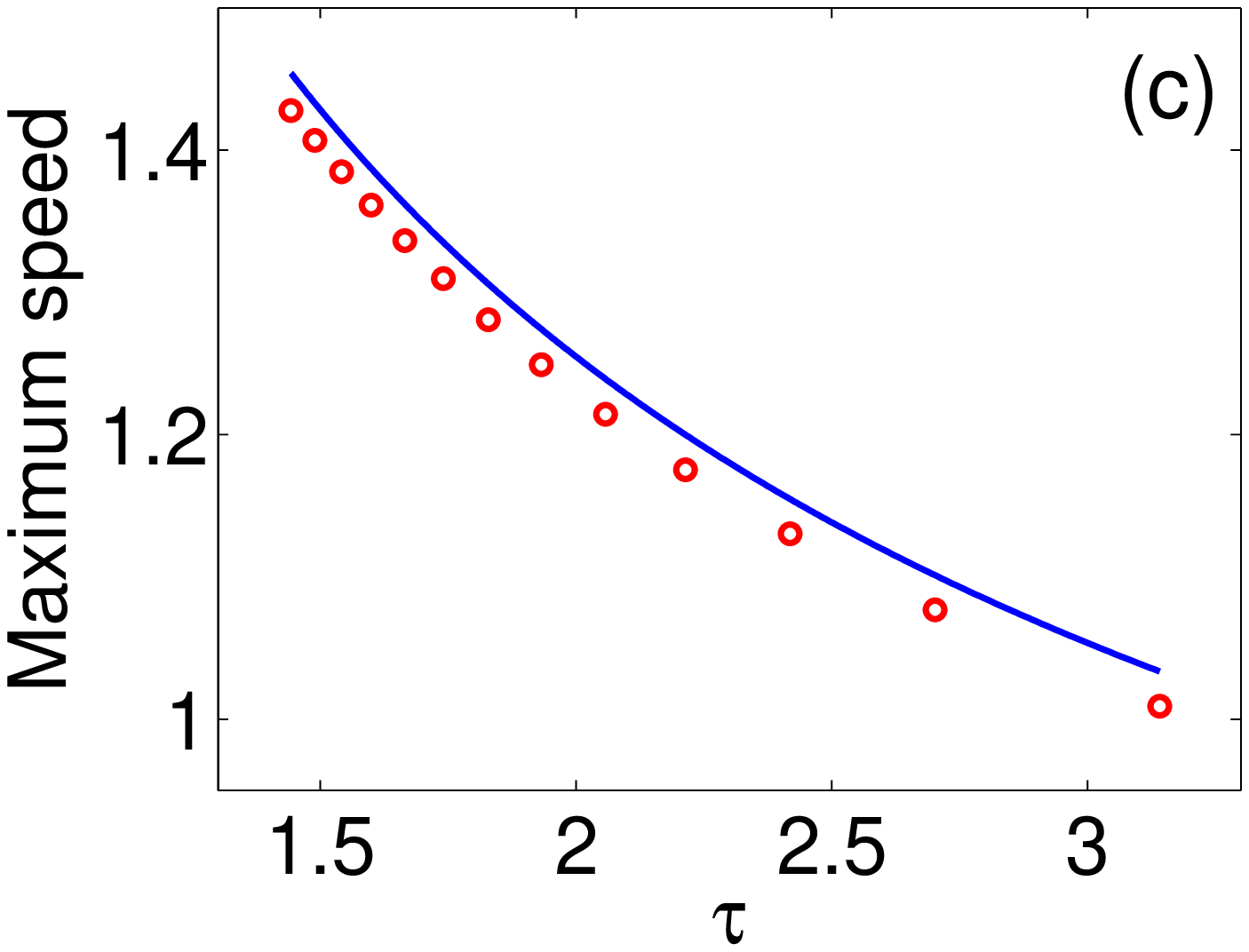} \label{Line_speed}}
\subfigure{\includegraphics[scale=0.26]{Circle_line_sim_map_a.eps} \label{Line_sim_map}}
\caption{Comparison of numerical simulation (red
  circular markers)  with the 
  analytical expressions (continuous blue curve) given by
    Eqs. \eqref{13_fourier_a}-\eqref{13_fourier_b} for (a) the amplitude, (b) period, and (c) the
    maximum speed of the   collapsed straight line orbit.  (d) At each value of  $a$, the time delay was chosen as $\tau =
\frac{2}{\sqrt{2a
    -1}}\arctan\left(\frac{\sqrt{2a-1}}{1-a}\right)$ (black circular
markers) to
  ensure asymptotic time convergence to the collapsed-straight line orbit state.}\label{Line_orbit}
\end{center}
\end{figure}

Figures \ref{Line_amplitude}-\ref{Line_sim_map} show a comparison between our
analytical results  given by Eqs. \eqref{13_fourier_a}-\eqref{13_fourier_b}  and results obtained using numerical simulation for the amplitude, period and maximum speed of
oscillation for different values of $a$ and $\tau$. There is excellent agreement in both amplitude and period between our analysis
and the numerical simulations [Figs. \ref{Line_amplitude}-\ref{Line_omega}]. The agreement for the speed of motion is
very good as well, but the theoretical
estimate is shifted slightly with respect to the results from
simulations [Fig. \ref{Line_speed}]. As in the collapsed circular orbit, we
note that the collapsed set of particles have a maximum speed which exceeds
one, the speed that individual, uncoupled particles acquire in the long-time
limit. As before, this effect arises from the attraction that the current particle position $\mathbf{R}(t)$ feels towards the delayed position $\mathbf{R}(t-\tau)$ when the latter lies in the direction of motion of the collapsed particles.

\section{The Effects of Noise on the Swarm}\label{sec:Noise}

In the absence of noise, the initial alignment
of the swarm particles is critical in determining the asymptotic behavior of
the swarm  (Sec.~\ref{sec:Comp}). When noise is introduced, the interplay of coupling strength, time
delay and noise intensity gives rise to very interesting behavior due to
fluctuations in the particles' alignment. Specifically, our studies show that if the coupling strength $a$ and/or the
time delay $\tau$ are below a certain limit, then the presence of noise
promotes swarm transitions from aligned into misaligned coherent states. More
surprising, however, is that if the coupling strength $a$ and/or the
time delay $\tau$ are big enough, then there is a noise intensity
threshold that forces a transition in the swarm from misaligned into aligned
states. In addition, we show that for these high values of $a$ and/or $\tau$,
 the system presents an interesting hysteresis phenomenon when the noise
intensity is time dependent.

For the purpose of these studies, we define the alignment of particle $j$ with the rest of the swarm as the cosine of the angle between
the velocity of particle $j$ and the velocity of the swarm as a whole:
\begin{align}
\cos\theta_j = \frac{\dot{\mathbf{r}}_j \cdot \dot{\mathbf{R}}}{|\dot{\mathbf{r}}_j| |\dot{\mathbf{R}}|}.
\end{align}
Therefore the alignment of individual particles ranges from -1
to 1. A good measure of the overall alignment of the swarm is furnished by
the ensemble average of these cosines given as
\begin{align}
\textrm{Mean swarm alignment} =  \frac{1}{N}\sum_{j=1}^N\cos\theta_j = \frac{1}{N}\sum_{j=1}^N\frac{\dot{\mathbf{r}}_j \cdot \dot{\mathbf{R}}}{|\dot{\mathbf{r}}_j| |\dot{\mathbf{R}}|}.
\end{align}

\begin{figure}[t!]
\begin{center}
\subfigure{\includegraphics[scale=0.26]{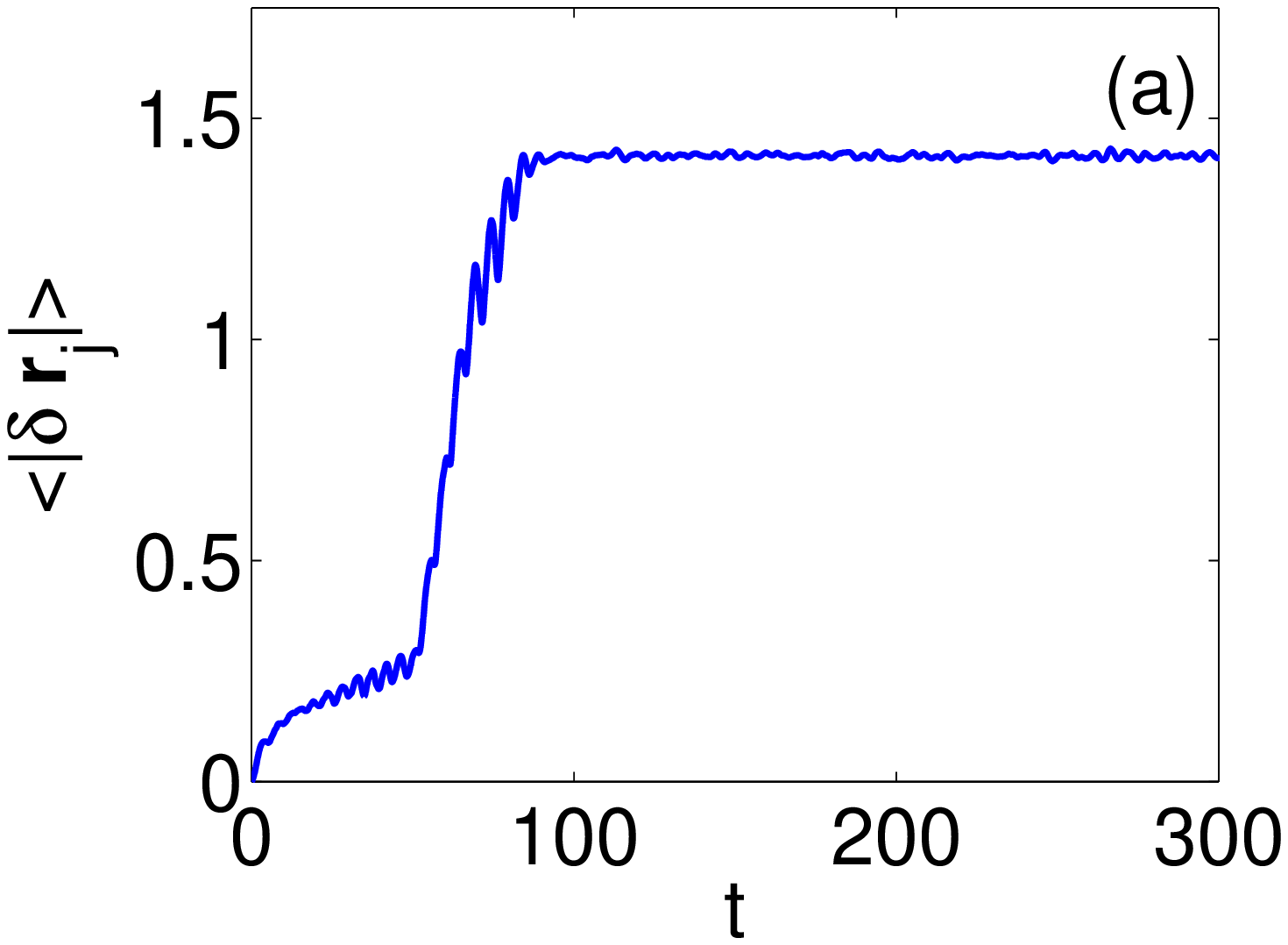} \label{dr_a_0p5_tau_0p5}}
\subfigure{\includegraphics[scale=0.26]{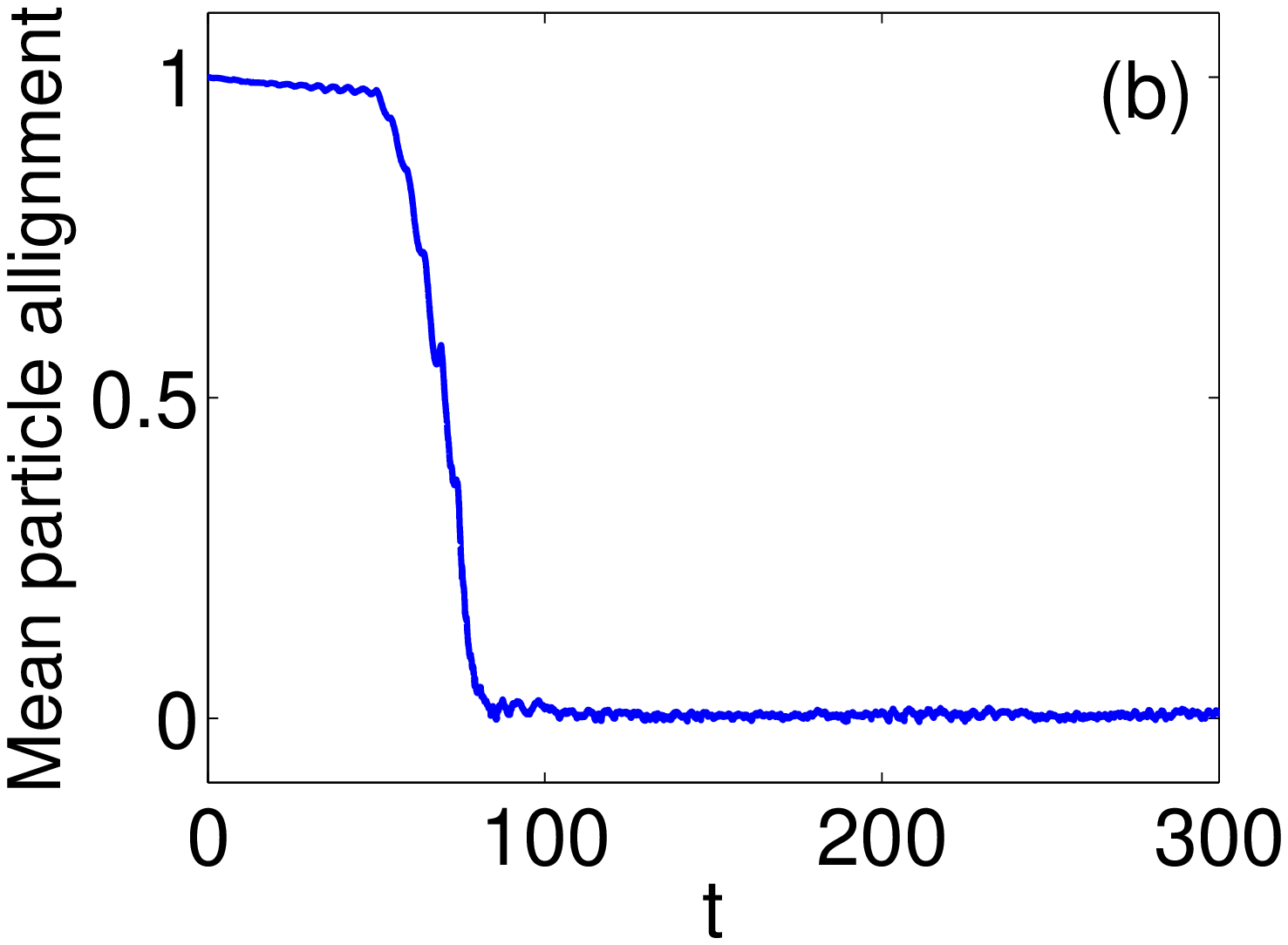} \label{allign_a_0p5_tau_0p5}}
\caption{ Time evolution of the ensemble average of (a) the particle distance to
    the center of mass, and (b) the mean particle alignment showing how the particle alignment breaks up
  due to the effects of noise. For long times the swarm converges to a ring
   state.  The parameter values  of $a = 0.5$ and
  $\tau = 0.5$ are associated with  region A of
Fig. \ref{a_tau_regions}. The time delay is
  turned on at $t=50$ and the noise standard deviation is $\sigma = 0.05$ ($D = 0.00125$). }\label{dr_allign_A}
\end{center}
\end{figure}

We first carry out a numerical simulation with coupling
constant $a=0.5$ and noise standard deviation $\sigma
= 0.05$ (noise intensity $D = 0.00125$).  At $t = 50$, a time delay of $\tau =
0.5$ is turned on. These parameters correspond to region A of
Fig. \ref{a_tau_regions}. Initially, we place all particles at the origin and
align their velocities by choosing $\dot{x}_j = 1$ and $\dot{y}_j = 1$ for
all particles. We describe the behavior of the swarm by following the ensemble averages of the particle distances to the center of mass
[Fig. \ref{dr_a_0p5_tau_0p5}] and of the particle
alignment [Fig. \ref{allign_a_0p5_tau_0p5}] as functions of time. Before the time delay is turned on at $t=50$,
the swarm is in a translating state with particles slightly spread out from
the center of mass in a `pancake' shape, as described in \cite{Erdmann05},
with an ensemble alignment close to one. Once the delay is turned on, the
translating state is broken up and the swarm converges to the ring state in
which the mean particle alignment is near zero. The radius of the ring
obtained in this numerical simulation matches the theoretical
 result [Eq. \eqref{ring_rho_omega}] that predicts a radius of $\frac{1}{\sqrt{a}} =
\sqrt{2} \approx 1.41$. A completely analogous situation ensues for parameters
in region B of Fig. \ref{a_tau_regions} (results not shown). In addition, in
both cases the swarm will immediately converge to the ring state if the
swarm velocities are not sufficiently aligned at time zero. We thus
conclude that for these choices of $(a, \ \tau)$ pairs, the noise misaligns
the particles' velocities and forces a transition into the ring state.

In contrast to the cases discussed above, for parameters in region C of
Fig. \ref{a_tau_regions}, a sufficiently large noise intensity promotes
transitions from misaligned to aligned states. We
show this by comparing the results of a series of simulations for different
values of the noise standard deviation $\sigma$. The simulations are divided
into two cases that differ only on the initial conditions for the swarm
particles. In all simulations, the coupling constant  $a=2$ and a time delay
of $\tau = 2$ is turned on at $t = 50$. In the first case, all particles start from the origin with identical velocities $\dot{x}_j = 1$
and $\dot{y}_j = 1$. In the second case, all swarm particles
are initially distributed uniformly on the unit square and
 are at rest.

\begin{figure}[t!]
\begin{center}
\subfigure{\includegraphics[scale=0.26]{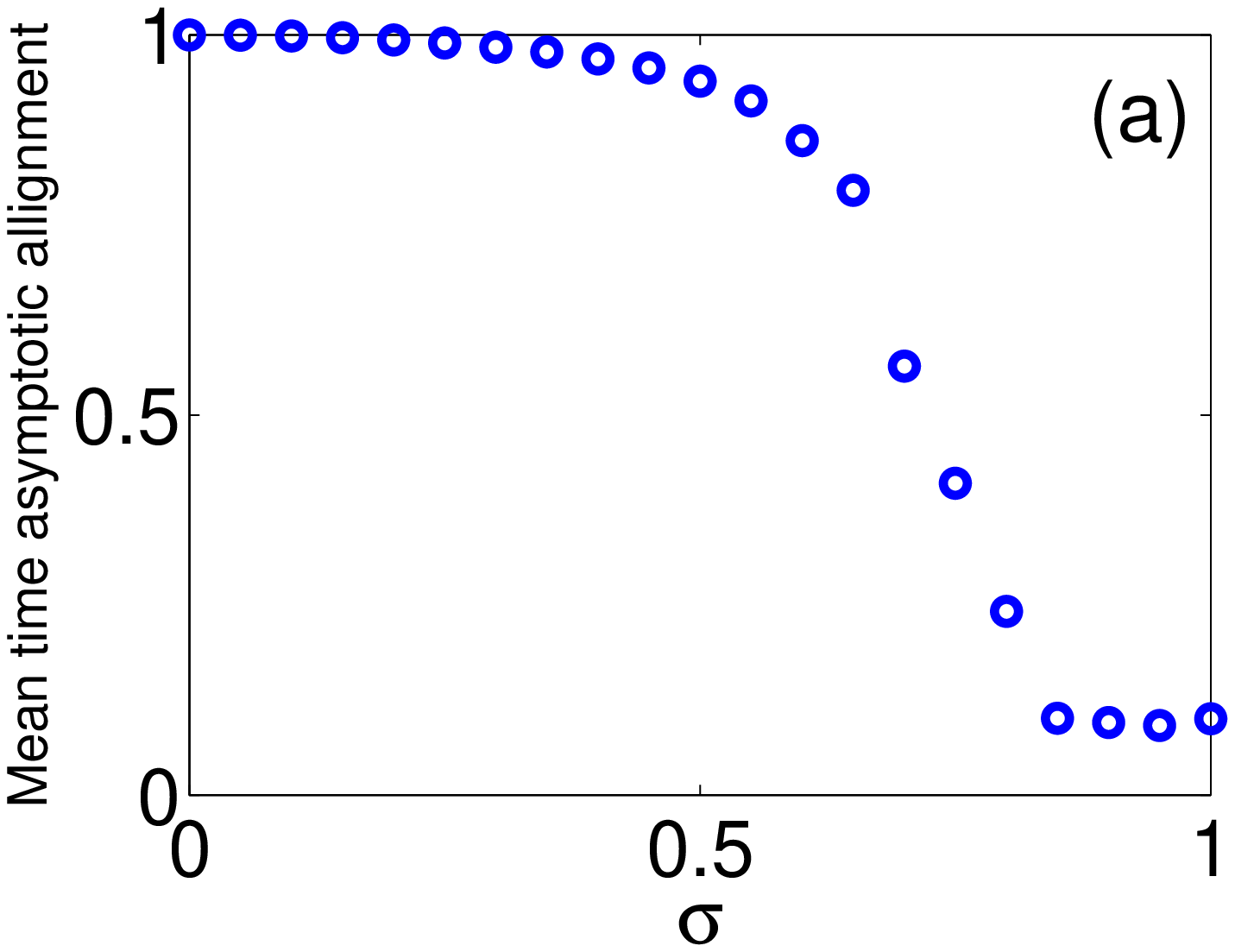} \label{asympt_allign}}
\subfigure{\includegraphics[scale=0.26]{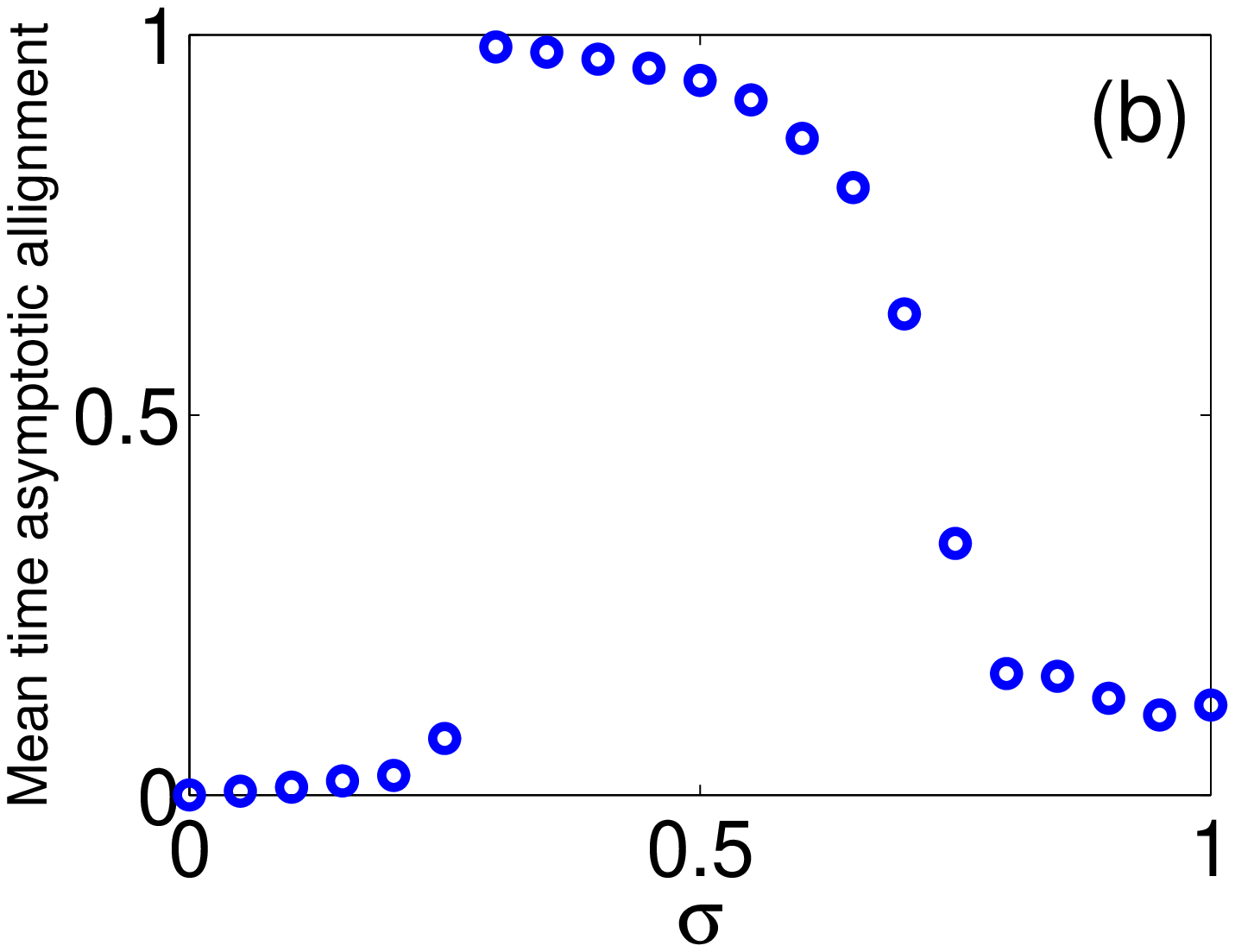} \label{asympt_unallign}}
\caption{Asymptotic value of the mean particle alignment for (a) particles
  starting with perfectly aligned velocities at time zero and
  (b) for particles distributed uniformly over the unit square and starting from rest for different values of the noise standard deviation $\sigma$. The
  parameter
  values of $a = 2$ and $\tau = 2$ (turned on at $t=50$) are
    associated with a location in region C of
Fig. \ref{a_tau_regions}. }\label{asympt_vel_proj}
\end{center}
\end{figure}

\begin{figure}[h!]
\begin{center}
\subfigure{\includegraphics[scale=0.26]{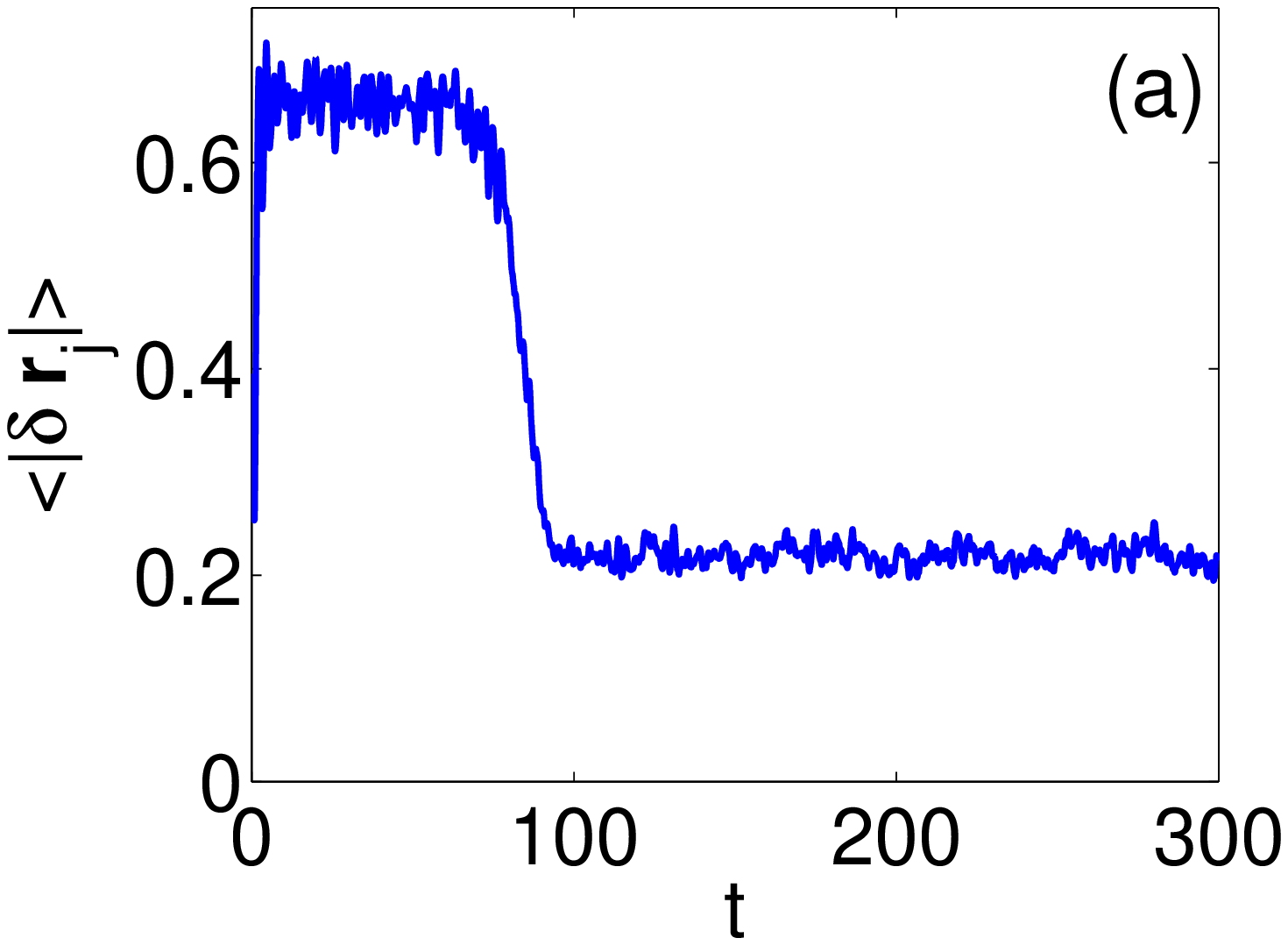} \label{dr_a_2_tau_2}}
\subfigure{\includegraphics[scale=0.26]{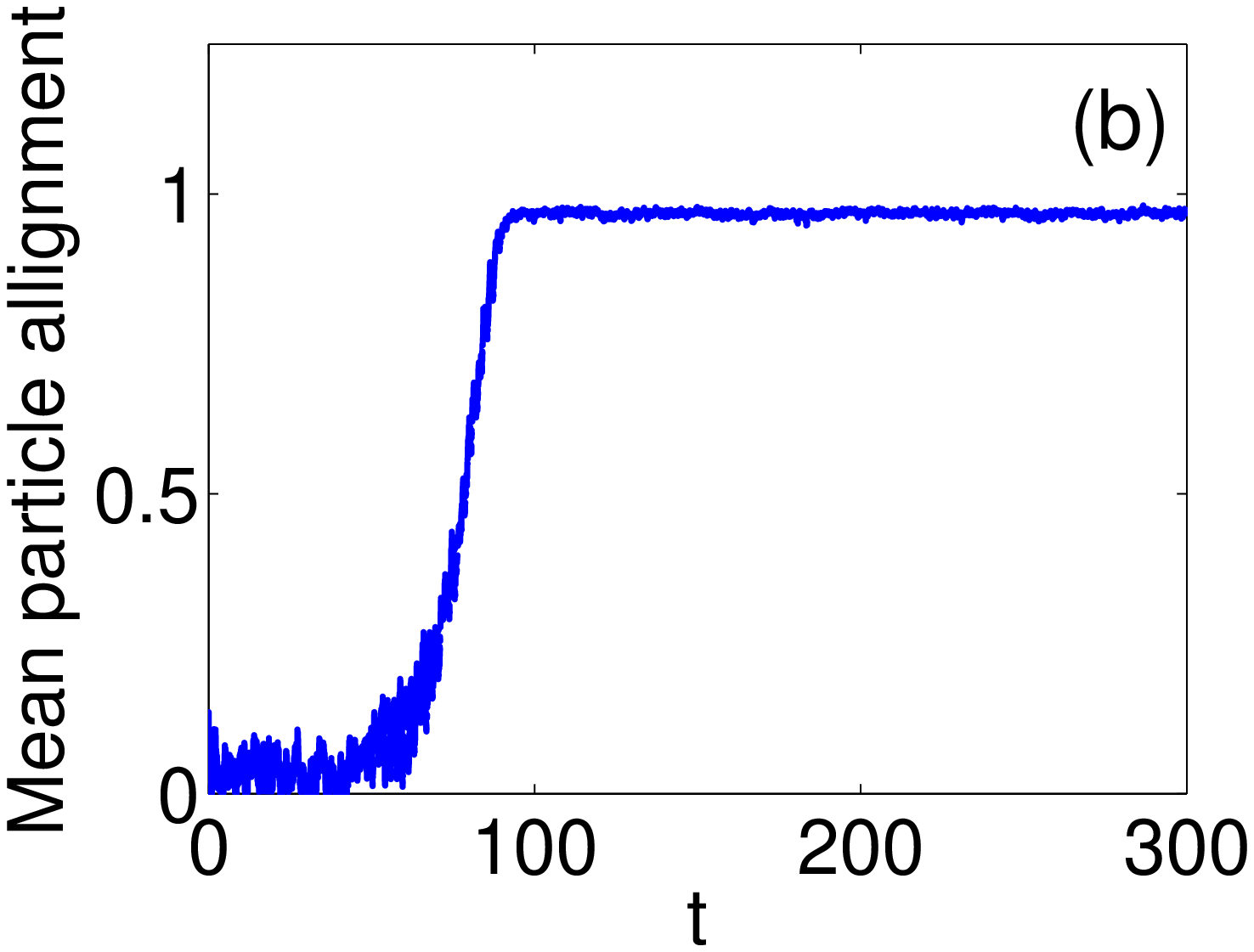} \label{allign_a_2_tau_2}}
\caption{ Time evolution of the ensemble average of (a) the particle distance to
    the center of mass, and (b) the mean particle alignment showing how the swarm transitions from a
  ring state into a compact, rotational state with alignment
  close to one. The parameter values  of $a = 2$ and $\tau = 2$ (turned on at
  $t=50$) and $\sigma = 0.4$ ($D = 0.08$) are associated with region C of
Fig. \ref{a_tau_regions}. Particles are initially distributed uniformly over
the unit square and start from rest.}\label{dr_allign_a_2_tau_2}
\end{center}
\end{figure}

In these simulations, the final state of the swarm may be visualized by plotting the mean swarm
alignment after transients have decayed ($t=300$) as a function of noise
intensity for the first case
[Fig. \ref{asympt_allign}] and the second case [Fig. \ref{asympt_unallign}].  In the first case of simulations, the high initial alignment of
particles' velocities forces the swarm to converge to a compact rotating state
independent of noise intensity. However, the rotational state is destroyed
if the noise standard deviation is bigger than $\sigma \approx 0.8$
[Fig. \ref{asympt_allign}]. The situation is more interesting and complex for the second set of
simulations. For low noise intensities ($\sigma \lesssim 0.26$) the low
initial alignment of the particles leads the swarm to converge to a ring
state with near zero mean alignment [Fig. \ref{asympt_unallign}]. A noise
standard deviation just beyond the threshold of $\sigma \approx 0.26$ displays an
interesting effect. 
As the $\sigma \approx 0.26$ threshold is crossed,
  the swarm transitions from the ring state into the rotating state with high mean alignment. An examination
of the full simulation data reveals that the transition occurs as
an increasing group of particles gradually becomes aligned and eventually absorbs
all the remaining particles. A sufficient amount of noise is necessary for
this transition, since it allows each particles' velocity vector to probe many
directions until finally enough of them become trapped in a `potential
well' of alignment with other particles. As with the first case of simulations a noise
standard deviation bigger than $\sigma \approx 0.8$ breaks up the rotating state. Figure \ref{dr_allign_a_2_tau_2} clearly shows the transition from the ring to
the compact, rotational state through the time evolution of the ensemble averages of the particle
distances to the center of mass and of the mean particle alignment.

\begin{figure}[t!]
\begin{center}
\subfigure{\includegraphics[scale=0.26]{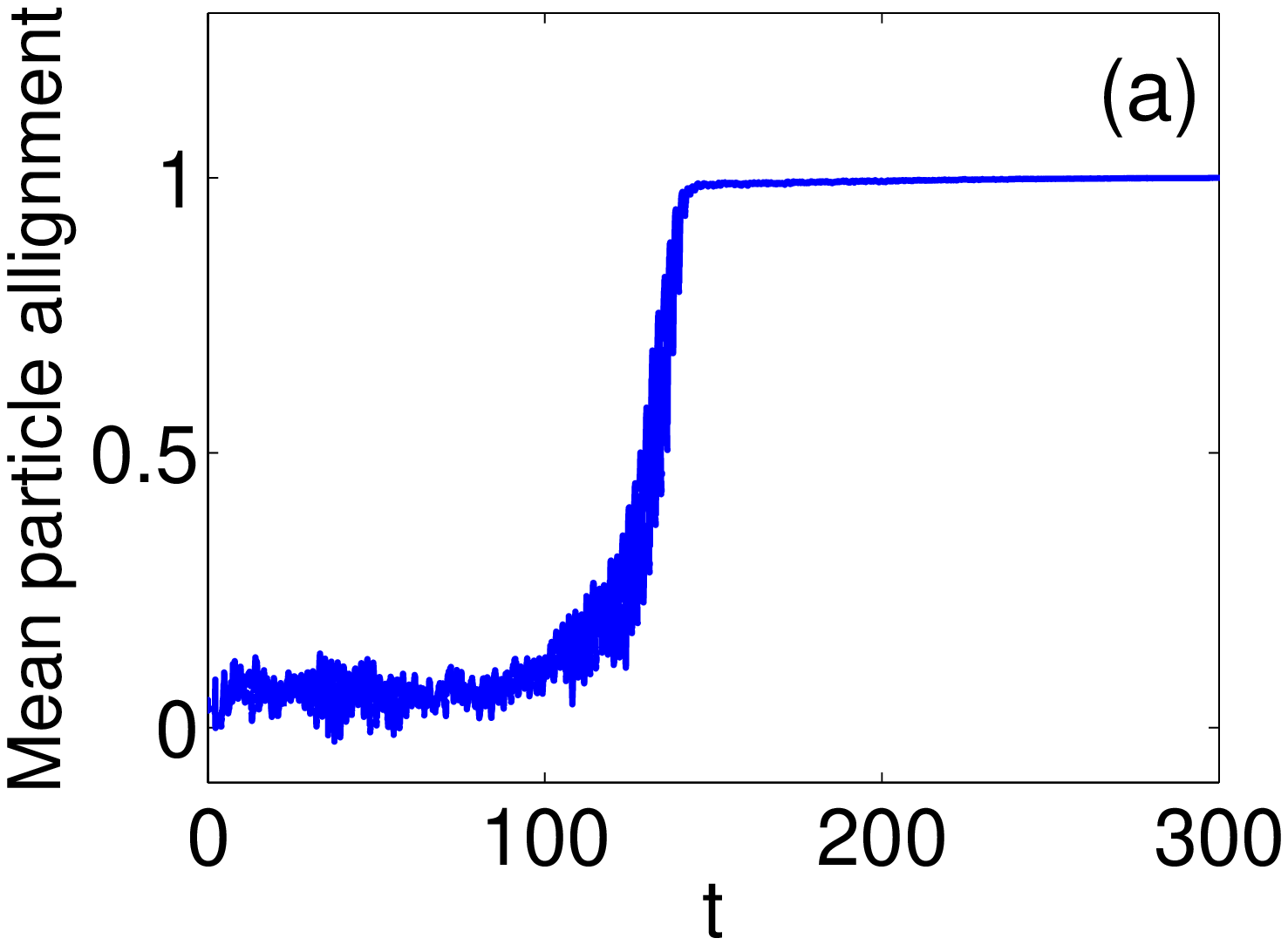} \label{allign_a_2_tau_2_hyst1}}
\subfigure{\includegraphics[scale=0.26]{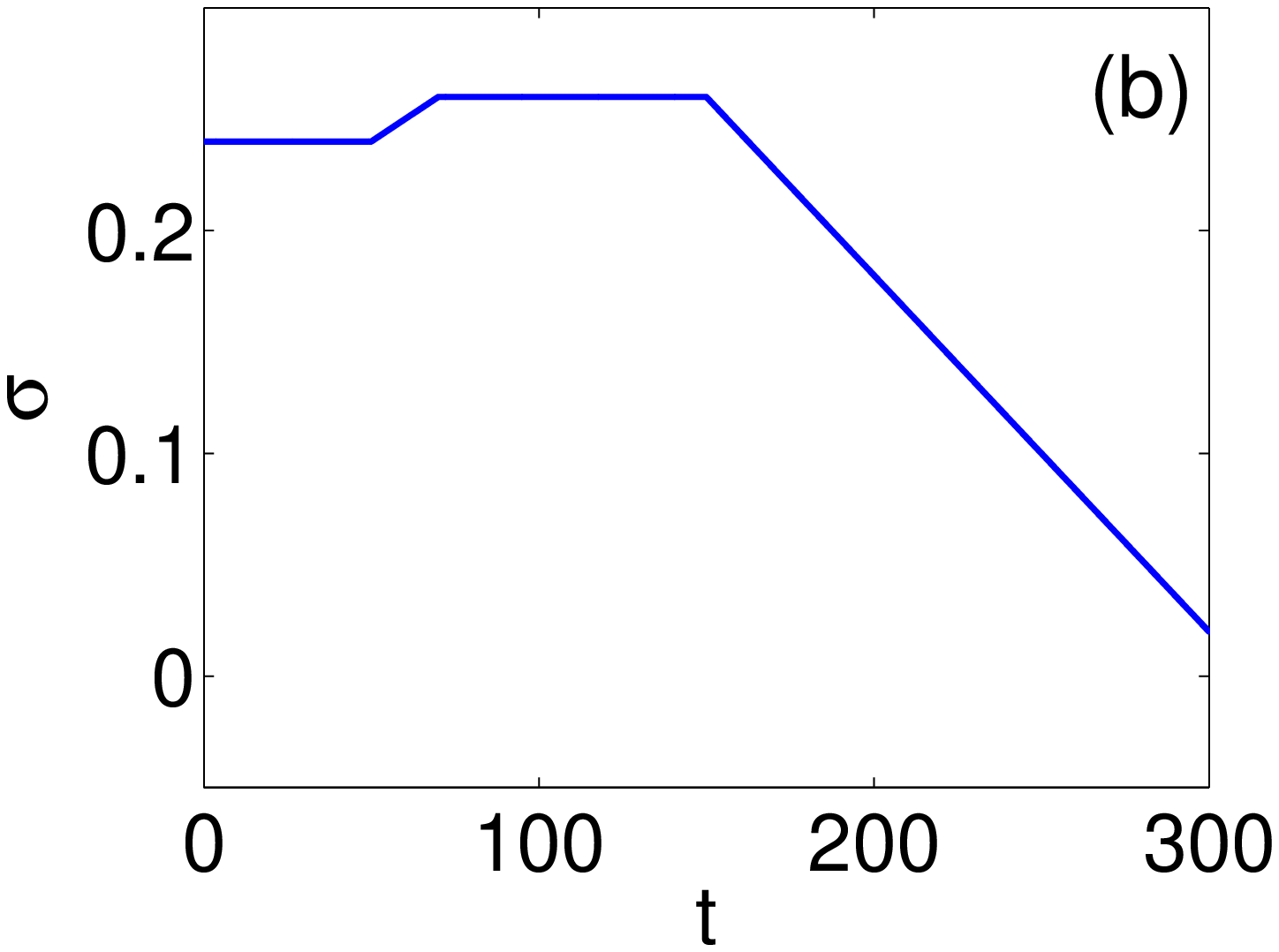} \label{sigma_a_2_tau_2_hyst1}}
\subfigure{\includegraphics[scale=0.26]{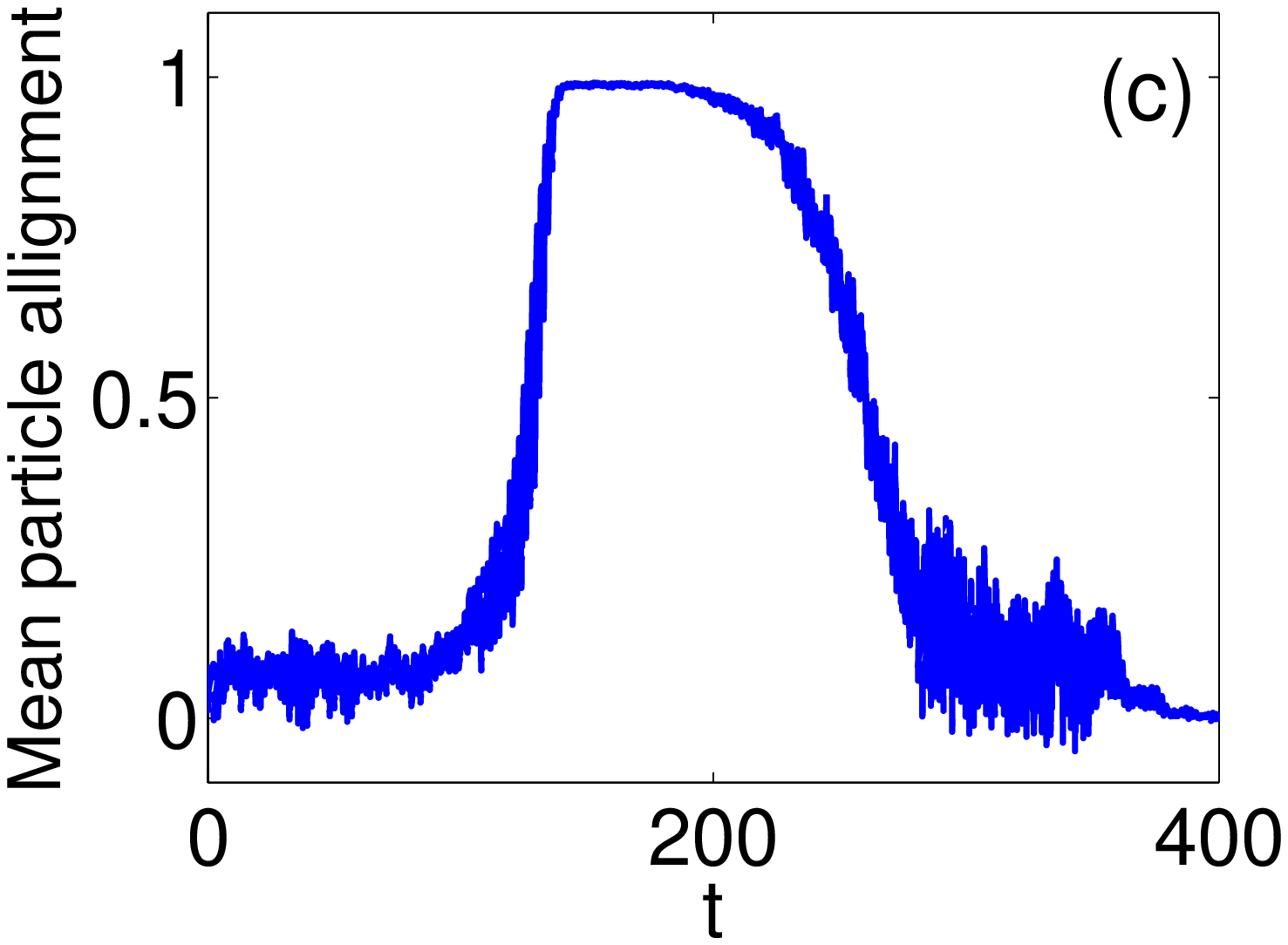} \label{allign_a_2_tau_2_hyst2}}
\subfigure{\includegraphics[scale=0.26]{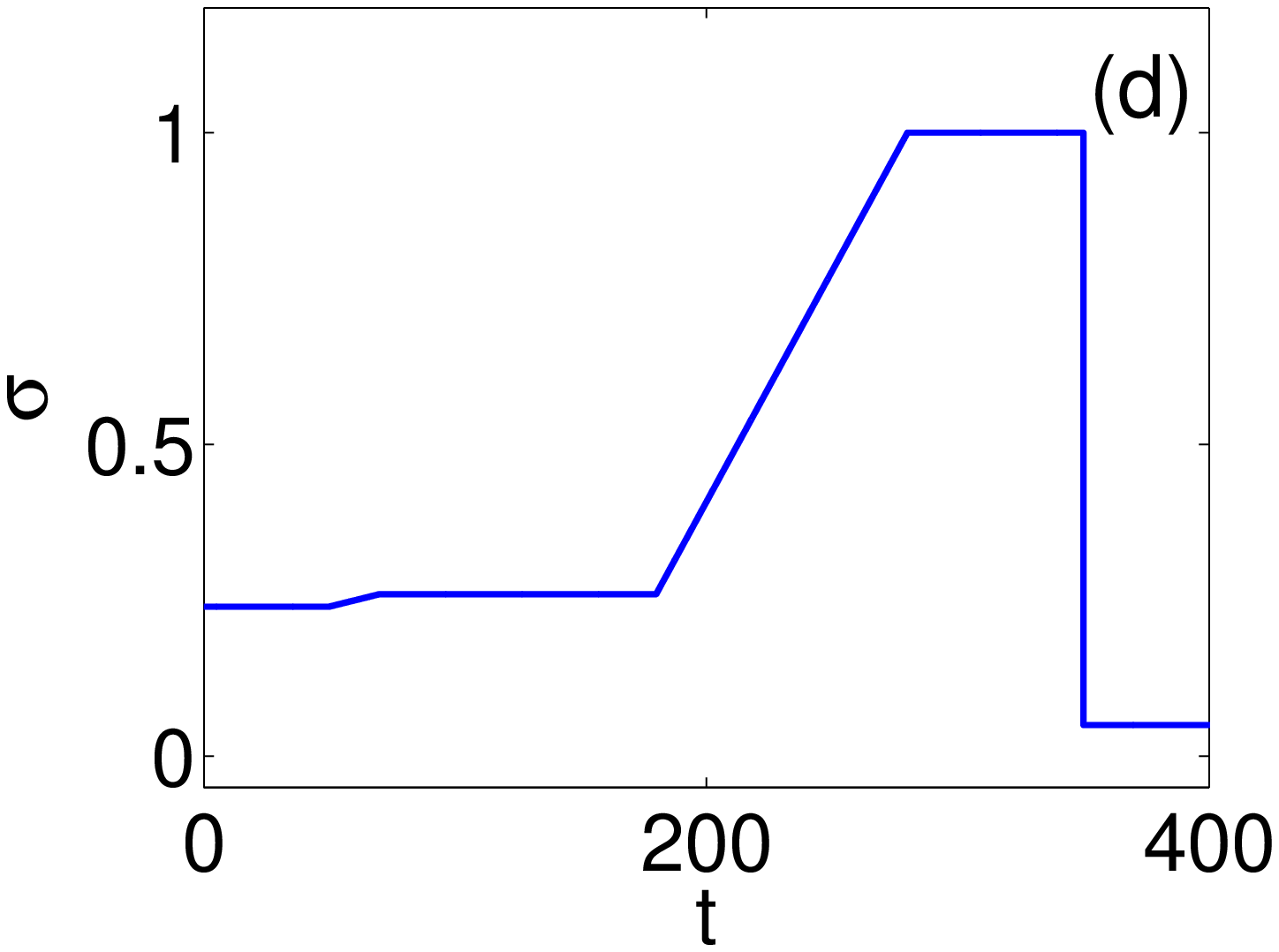} \label{sigma_a_2_tau_2_hyst2}}
\caption{Time evolution of (a) mean particle alignment for example 1, (b)
  noise standard deviation for example 1, (c) mean particle alignment for
  example 2, and (d) noise standard deviation for example 2.  The results
  show how a time-dependent noise
  intensity may be used to force swarm transitions.  The parameter values
  of $a = 2$ and $\tau = 2$ (turned on at
  $t=10$)  are associated with region C of
Fig. \ref{a_tau_regions}. Particles are initially distributed uniformly over
the unit square and start from rest.}\label{hyst}
\end{center}
\end{figure}

Further studies on the switching behavior between coherent states
of the swarm demonstrate that the system exhibits a hysteresis phenomena. With the swarm system starting on the ring state with noise standard deviation
of $\sigma = 0.24$, one  can force a transition into the rotating state by
increasing the noise to $\sigma = 0.26$. However, even if the noise is lowered
down to $\sigma = 0.02$, the swarm remains in the rotating state with a high
velocity alignment [Figs. \ref{allign_a_2_tau_2_hyst1}-\ref{sigma_a_2_tau_2_hyst1}]. Nevertheless, it is possible to return the swarm to the ring state if, once in
the rotating state, the
noise is raised to very high amounts ($\sigma = 1$) for a sufficient amount of
time and then dropped suddenly to a very low value ($\sigma = 0.05$). The high noise levels serve to completely misalign the
particles' velocities and allow them to converge to the ring once the noise
levels are below $\sigma \lesssim 0.26$.

\section{Conclusions}\label{sec:Conc}

In this work we analyzed the dynamics of a self-propelling swarm where
  individuals interact with a
communication time delay  in the presence of noise. Using a mean field
approximation in the  deterministic case, we analytically obtained the
 complete bifurcation picture in the parameter space of coupling strength and
communication time delay. This analysis shows how different combinations of coupling
strength and time delay induce the swarm to adopt different coherent
structures asymptotically in time. Our bifurcation studies demonstrated
the existence of a Bogdanov-Takens point, where the stationary center of mass
solution has a double zero eigenvalue, which is critical in organizing the
dynamics of the swarm.

{The stable patterns that are possible for this system have several
applications for autonomous vehicles. More detailed applications for each
pattern are as follows: (1) the translational state may be used for target tracking and group transport \cite{Morgan05,chung2006}. (2) The ring
state should prove useful in terrain coverage and regional surveillance \cite{Svennebring03,
  Vallejo09}. (3) The rotating state may be exploited in obstacle avoidance,
boundary tracking and surveillance \cite{hsieh2005, Morgan05, Vallejo09}.
In addition, we believe all three patterns are applicable to the problem of environmental sensing \cite{lynch2008,lu2011}.}

In numerical experiments with noise, we showed that the interplay of coupling
strength, time delay and noise intensity may give rise to interesting
switching behavior from one coherent structure to another. We found that if the coupling strength $a$ and/or the time delay $\tau$ are
below a certain limit, then the presence of noise induces transitions from
states in which the alignment of the particles' velocities is high into states
with low alignment. More surprising, however, is that if the coupling strength $a$ and/or the
time delay $\tau$ are big enough, then there is a noise intensity
threshold that forces a transition in the swarm from misaligned into aligned
states. In addition, by using a time-dependent noise intensity at these high values of
$a$ and/or $\tau$, we show that the system exhibits hysteresis since the
swarm's transitions are not easily reversible. {We note that analytical
  results on the effects of noise on delay-coupled swarms are not easy to  obtain. Two
  examples relevant to our work are given in \cite{Erdmann05,Strefler08},
  where the authors investigate models similar to the one presented here but without time delay.}

Realistic application of the model treated here to the motion of multi-robot
systems requires local repulsion among individuals to be taken into
account. We have simulated the swarm model with the addition of a
repulsive inter-agent potential of exponential form $U_{ij} = c_r
\exp{\left({-\frac{|\mathbf{r}_i - \mathbf{r}_j|}{L_r}}\right)}$. These simulations demonstrate (results not shown) that
the coherent patterns we discussed in this article persist when the
characteristic repulsion length $L_r$ and repulsion strength $c_r$ between
robots are small compared to global attraction parameters. Stronger repulsion can destabilize the coherent structures.

{Recently,  systems with non-uniform time delays have received much
  attention. For example, the important question of synchronization in networks
  communicating at randomly-distributed time delays has been recently investigated \cite{Masoller05, Masoller06}. In practical applications, the case of differing (and even time-varying)
  time delays between agents may be treated similarly to the case of a single
  delay by using a data buffer \cite{Yang10}. The idea is to identify an upper
  bound to the time delay ($\tau_\textrm{max}$) between all agent pairs and
  then design the agents so that the actuation occurs when the data buffer of
  size $\tau_\textrm{max}$ is full.}

{As part of our ongoing work, we are extending our investigations for the cases
in which: (\emph{i}) the communication time delays vary between different
pairs of agents; and (\emph{ii}) the communication graph is non-globally
coupled. In realistic settings, both of these cases may occur due to the effects
of the spatial distribution of agents such as signal travel times and
imperfect transmission arising, for example, from  complex terrain topography
or component malfunction. In the case of communication delays that differ
among different pairs of agents (though constant in time), our preliminary results show some patterns
analogous to the ones observed here, but with much more added complexity. The present investigation lays a good foundation on which to base the study of these more complicated cases.}

In summary, our results aid in understanding the stability of complex coherent
structures in swarming systems with time delayed communication and in the
presence of a noisy environment. Although our analytical and numerical results were
obtained using a model with linear, attractive interactions, our analysis
gives useful insight for the study of models with more general forms of
time delayed coupling between agents. Our results may prove to be valuable for
the control of man-made vehicles where actuation and communication are delayed, as well as in
understanding swarm alignment in biological systems.  

\appendices

\section{Analysis of the Ring State}\label{ring}

The swarm ring state is obtained when the center of mass is stationary. For the solution $\mathbf{R}=$const. to satisfy Eq. \eqref{CM} we require
\begin{align}\label{dri_condition}
\sum_{i=1}^N\delta\dot{\mathbf{r}}_i^2 \delta\dot{\mathbf{r}_i} = 0. 
\end{align}

We simplify Eq. \eqref{dri} by taking $\mathbf{R}=$const. and using
Eq. \eqref{dri_condition} { we obtain}
\begin{align}\label{dri_ring}
\delta\ddot{\mathbf{r}}_j=&
\left(1 -\delta\dot{\mathbf{r}}_j^2\right)\delta\dot{\mathbf{r}}_j - a  \delta\mathbf{r}_j - \frac{a}{N}\delta\mathbf{r}_j(t-\tau).
\end{align}
We consider the  {large system size} limit $N\rightarrow \infty$ and we drop the
delayed term. The resulting equations are simply ODEs and so the analysis
below shows that the ring orbit is not dependent on having time delays in the
system. Writing Eq. \eqref{dri_ring} in polar coordinates $\delta x_j = \rho_j
\cos{\theta_j}$ and  $\delta y_j =
\rho_j \cos{\theta_j}$, we obtain
\begin{subequations}
\begin{align}
\ddot{\rho}_j   =& \left(1 - \dot{\rho}_j^2 - \rho_j^2 \dot{\theta}_j^2 \right)\dot{\rho}_j + \rho_j \dot{\theta}_j^2 - a \rho_j,\label{rho_theta_a}\\
\rho_j \ddot{\theta}_j =& \left(1 - \dot{\rho}_j^2 - \rho_j^2 \dot{\theta}_j^2\right) \rho_j \dot{\theta}_j - 2 \dot{\rho}_j \dot{\theta}_j.\label{rho_theta_b}
\end{align}
\end{subequations}

Equations \eqref{rho_theta_a}-\eqref{rho_theta_b} have the trivial solution $\rho_j = 0$ as well as a ring solution:
\begin{gather}
\rho_j = \frac{1}{\sqrt{a}}, \qquad \dot{\theta}_j =  \pm\sqrt{a},
\end{gather}
in which particles move at unit speed, $\rho_j \dot\theta_j = \pm 1$.

\section{Analysis of the Rotating State}\label{rotating}

In the noiseless rotating state, all particles collapse to a point,
$\delta\mathbf{r}_i=0$, and the equation for the center of mass given by
Eq. \eqref{CM} simplifies considerably to
\begin{align}\label{CM_circle}
\ddot{\mathbf{R}}=& \left(1 - \dot{\mathbf{R}}^2\right)\dot{\mathbf{R}} - a\left(\mathbf{R}(t) - \mathbf{R}(t-\tau)\right).
\end{align}
We write $\mathbf{R} = (X, Y)$ and introduce polar coordinates $X = \rho
\cos{\theta}$ and $Y = \rho \sin{\theta}$ to obtain 
\begin{subequations}\begin{align}
\ddot{\rho}   =& \left(1 - \dot{\rho}^2 - \rho^2 \dot{\theta}^2
\right)\dot{\rho} + \rho \dot{\theta}^2 - a \bigg(\rho - \rho_\tau\cos(\theta - \theta_\tau)\bigg),\label{rho_theta_CM_circle_a}\\
\rho \ddot{\theta} =& \left(1 - \dot{\rho}^2 - \rho^2 \dot{\theta}^2\right)
\rho \dot{\theta} - 2 \dot{\rho} \dot{\theta} + a \rho_\tau\sin(\theta - \theta_\tau),\label{rho_theta_CM_circle_b}
\end{align}
\end{subequations}
where we've written $\rho_\tau \equiv \rho(t-\tau)$ and $\theta_\tau \equiv
\theta(t-\tau)$. Equations \eqref{rho_theta_CM_circle_a}-\eqref{rho_theta_CM_circle_b}  have a circular orbit solution, $\rho =
\rho_0$ and $\theta = \omega t + \theta_0$, where
\begin{subequations}
\begin{align}
\omega^2 =& a \cdot(1 - \cos\omega\tau),\label{omega_CM_circle_app_a}\\
\rho_0 =& \frac{1}{|\omega|} \sqrt{1 - a\frac{\sin\omega\tau}{\omega}}.\label{rho_CM_circle_app_b}
\end{align}
\end{subequations}
and $\theta_0$ is obtained from the initial conditions. In the main text we discuss the behavior of the solutions to Eqs. \eqref{omega_CM_circle_app_a}-\eqref{rho_CM_circle_app_b}.

\section{Analysis of the Degenerate Rotating State}\label{deg_rotating}

When the motion of the whole swarm is initially constrained to a line,
Eqs. \eqref{swarm_eq_a}-\eqref{swarm_eq_b} dictate that the swarm will remain on this line for all
times. If the coupling parameter $a$ and/or the time delay $\tau$ are large
enough, the resulting motion is a degenerate form of the rotating solution in which the swarm moves back and forth along a straight line.

In the case without noise all particles collapse to a point, $\delta\mathbf{r}_i=0$, and
the line along which motion occurs is arbitrary; here we use $X = Y$. The
problem reduces to analyzing a single delay equation { given by}
\begin{align}\label{CM_line}
\ddot{X}=& \left(1 - 2\dot{X}^2\right)\dot{X} - a\left(X(t) -
  X(t-\tau)\right).
\end{align}
We  {find} a solution using Fourier analysis. {We let}
\begin{align}\label{X_fourier}
X(t) = \sum_{n = -\infty}^{\infty} c_n e^{in\omega t},
\end{align}
where the coefficients satisfy $c_n = {c_{-n}}^*$ in order to ensure that
$X(t)$ is a
real quantity. Substituting Eq. \eqref{X_fourier} into Eq. \eqref{CM_line}, we
get for the $n$-th mode
\begin{align}\label{n_fourier}
  - n^2 \omega^2 c_n  &= i  n \omega c_n \notag\\
&+ 2i \omega^3 \sum_{\ell, m \neq 0} c_\ell c_m
c_{n-\ell-m}\ell m (n - \ell - m) \notag\\
&- a c_n (1 - e^{-in\omega \tau}),         
\end{align}
for $n = 0,1,2,\dots$. The $n=0$ equation is
\begin{align}\label{0_fourier}                                                   
\sum_{\ell, m\neq 0} c_\ell c_m c_{-\ell-m} \ell m (\ell + m) = 0,
\end{align}
which does not involve $c_0$. Unsurprisingly, $c_0$ is undetermined since the position of the center of mass may be translated in space without modifying the dynamics of the system.

We now approximate the motion of the center of mass by keeping the first three
modes. By appropriately choosing the time origin, we may take $c_1$ to be
purely real and positive. In contrast, $c_2$ and $c_3$ are complex quantities
which we write as $c_i = |c_i| e^{i\phi_i}$, for $i=2,3$. The equations for
the first three modes $n=1,2,3$ become
\begin{subequations}
\begin{align}
-& \omega^2 c_1  = i \omega c_1 \notag\\
&+ 2i \omega^3\left( -3c_1^3 - 36 c_2^2c_3^* -
  54c_1|c_3|^2 - 24c_1|c_2|^2 + 9c_1^2c_3 \right)\notag\\
 &- a c_1(1 - e^{-i\omega\tau}),\label{123_fourier_a}\\
\notag\\
-& 4\omega^2 c_2  = 2i \omega c_2 \notag\\
&+ 2i \omega^3\left( -108c_2|c_3|^2 - 36
  c_1c_2^* c_3 - 24c_2|c_2|^2 - 12c_2c_1^2 \right)\notag\\           
 &- a c_2(1 - e^{-2i\omega\tau}),\label{123_fourier_b}\\
\notag\\
-& 9\omega^2 c_3  = 3i \omega c_3 \notag\\
&+ 2i \omega^3\left( -18c_3c_1 -
  72c_3|c_2|^2- 81c_3|c_3|^2 - 12c_2^2c_1 + c_1^3 \right)\notag\\              
 &- a c_3(1 - e^{-3i\omega\tau}).\label{123_fourier_c}
\end{align}
\end{subequations}

In addition, the condition from Eq.\eqref{0_fourier} becomes
\begin{align}\label{0_fourier_bis}
6(c_2 c_3^* - c_2^*c_3) - c_1(c_2 - c_2^*)=0.
\end{align}
Separating Eqs. \eqref{123_fourier_a}-\eqref{123_fourier_c} and Eq. \eqref{0_fourier_bis} into real and
imaginary parts yields a system of seven equations (since the real part of
Eq. \eqref{0_fourier_bis} is satisfied automatically) for the six unknowns:
$\omega$, $c_1$, $|c_2|$, $\phi_2$, $|c_3|$ and $\phi_3$. These equations cannot be
satisfied in general. However, if $|c_2| = 0$, then the equation for mode
$n=2$ [Eq. \eqref{123_fourier_b}] and Eq. \eqref{0_fourier_bis} are satisfied automatically,
leaving
four equations:
\begin{subequations}
\begin{align}                                                                   
- \omega^2 c_1  =& i \omega c_1 + 2i \omega^3\left( -3c_1^3 54c_1|c_3|^2 +
  9c_1^2c_3 \right) \notag\\
&- a c_1(1 - e^{-i\omega\tau}),\label{13_fourier_a}\\
- 9\omega^2 c_3  =& 3i \omega c_3 + 2i \omega^3\left( -18c_3c_1 - 81c_3|c_3|^2
  + c_1^3 \right) \notag\\
&- a c_3(1 - e^{-3i\omega\tau})\label{13_fourier_b}.                         
\end{align}
\end{subequations}
for the four unknowns $\omega$, $c_1$, $|c_3|$ and $\phi_3$. Equations \eqref{13_fourier_a}-\eqref{13_fourier_b}
may be solved numerically and permit one to approximate the motion of the
center of mass in the form
\begin{align}
X(t)=Y(t) = 2 c_1 \cos\omega t + 2|c_3| \cos(3\omega t + \phi_3).
\end{align}

The frequency of the straight line orbit of the swarm center of mass is
approximately equal to the frequency of the circular orbit in Eq. \eqref{omega_CM_circle}. In addition, the amplitude of oscillation of the straight line orbit is approximately equal to the radius of the circular orbit of Eq. \eqref{rho_CM_circle} divided by a factor of $\sqrt{6}$.

\section*{Acknowledgments}
The authors gratefully acknowledge the Office of Naval Research for its
support.  LMR and IBS are  supported by
Award Number R01GM090204 from the National Institute
Of General Medical Sciences. The content is solely the
responsibility of the authors and does not necessarily represent
the official views of the National Institute Of General
Medical Sciences or the National Institutes of Health. EF
is supported by the Naval Research Laboratory (Award No.
N0017310-2-C007).  We also extend our thanks to Kevin Lynch and M. Ani Hsieh for reading
early versions of the manuscript.


\clearpage

\end{document}